\documentclass[%
 reprint,
superscriptaddress,
 amsmath,amssymb,
 aps,prx,
]{revtex4-2}

\usepackage{graphicx}% Include figure files
\usepackage{dcolumn}% Align table columns on decimal point
\usepackage{bm}% bold math
\usepackage{hyperref}% add hypertext capabilities
\usepackage{physics}
\usepackage{xcolor}
\usepackage{soul}
\usepackage{subfigure}
\usepackage{bbold}
\usepackage{verbatim}
\usepackage{float}

\begin{document}

\title{Maxwell's demon across the quantum-to-classical transition}

\author{Björn Annby-Andersson}
 \email{bjorn.annby-andersson@teorfys.lu.se}
\affiliation{Physics Department and NanoLund, Lund University, Box 118, 22100 Lund, Sweden.}

\author{Debankur Bhattacharyya}
\affiliation{Institute for Physical Science and Technology, University of Maryland, College Park, Maryland 20742, USA.}

\author{Pharnam Bakhshinezhad}
\thanks{This author was formerly known as Faraj Bakhshinezhad.}
\affiliation{Atominstitut, Technische Universität Wien, Stadionallee 2, 1020 Vienna, Austria.}
\affiliation{Physics Department and NanoLund, Lund University, Box 118, 22100 Lund, Sweden.}

\author{Daniel Holst}
\affiliation{Physics Department and NanoLund, Lund University, Box 118, 22100 Lund, Sweden.}

\author{Guilherme De Sousa}
\affiliation{Department of Physics, University of Maryland, College Park, MD 20742, USA.}

\author{Christopher Jarzynski}
\affiliation{Institute for Physical Science and Technology, University of Maryland, College Park, Maryland 20742, USA.}

\author{Peter Samuelsson}
\affiliation{Physics Department and NanoLund, Lund University, Box 118, 22100 Lund, Sweden.}

\author{Patrick P. Potts}
\affiliation{Department of Physics and Swiss Nanoscience Institute,
University of Basel, Klingelbergstrasse 82, 4056 Basel, Switzerland.}

\date{\today}% It is always \today, today,

\begin{abstract}
In scenarios coined Maxwell's demon, information on microscopic degrees of freedom is used to seemingly violate the second law of thermodynamics. This has been studied in the classical as well as the quantum domain. In this paper, we study an implementation of Maxwell's demon that can operate in both domains. In particular, we investigate information-to-work conversion over the quantum-to-classical transition. The demon continuously measures the charge state of a double quantum dot, and uses this information to guide electrons against a voltage bias by tuning the on-site energies of the dots. Coherent tunneling between the dots allows for the buildup of quantum coherence in the system. Under strong measurements, the coherence is suppressed, and the system is well-described by a classical model. As the measurement strength is further increased, the Zeno effect prohibits interdot tunneling. A Zeno-like effect is also observed for weak measurements, where measurement errors lead to fluctuations in the on-site energies, dephasing the system. We anticipate similar behaviors in other quantum systems under continuous measurement and feedback control, making our results relevant for implementations in quantum technology and quantum control.
\end{abstract}

\maketitle

%\tableofcontents

\section{Introduction}

Maxwell's demon \cite{MaxwellTheory_of_heat-book,Maxwell-demon-2-book,Maruyama-RMP-2009} is the thought experiment where an external agent uses information acquired by a measurement in a feedback loop to rectify the energy flows of a physical system. Rectifying the flows of energy leads to a decrease of entropy in the studied system, seemingly violating the second law of thermodynamics. This paradox is resolved by accounting for information processing in the thermodynamic book-keeping, creating a link between information and energy \cite{Sagawa-Prog.Theor.Phys-2012,Parrondo-Nat.Phys-2015,Goold-J.Phys.A-2016}, as famously realized by Bennett \cite{bennett1982}, building on the results by Landauer \cite{landauer1961}. Over the last decades, the increase of experimental control over microscopic systems has fueled the development of stochastic \cite{review1,review2,review3,review4,review5,review6,Esposito-RMP-2009,review8,vandenBroeck-PA-2015,Peliti-book}, information \cite{Sagawa-Prog.Theor.Phys-2012,Parrondo-Nat.Phys-2015,Goold-J.Phys.A-2016}, as well as quantum \cite{Vinjanampathy-Contem.Phys-2016,QuantumThermoBook-Binder} thermodynamics, and resulted in numerous experimental implementations of the demon \cite{Serreli-Nature-2007,Toyabe-Nat.Phys-2010,Koski-PNAS-2014,Chida-Nat.Com.-2017,Kumar-Nature-2018,Ribezzi-Nat.Phys.-2019,Vidrighin-PRL-2016,Cottet-PNAS-2017,Masuyama-Nat.Com.-2018,Naghiloo-PRL-2018}.

An excellent platform for studying thermodynamics on the microscopic scale is provided by semiconductor quantum dots \cite{Pekola-Nat.Phys-2015,Hofmann-PSSb-2017,Josefsson-NatureNanotech-2018,Dorsch-NanoLetters-2021,Scandi-PRL-2022,Barker-PRL-2022,Haldar-2024}. These systems contain only a few electrons and exchange energy and electrons with their environment. Using a charge-sensing element, such as a quantum point contact or a single-electron transistor, it is possible to accurately detect and control single electrons in these structures in real time \cite{Lu-Nature-2003,Fujisawa-APL-2004,Vandersypen-APL-2004,Gustavsson-PRL-2006, Fujisawa-Science-2006,Utsumi-PRB-2010,Barker-PRL-2022}, making them well-suited for studying feedback control on the microscopic level. Two quantum dots in series, separated by a tunnel barrier, form a double quantum dot (DQD) \cite{vanderWiel-RevModPhys-2002}, featuring quantum effects such as superpositions of charge states and electron tunneling. This tunnel barrier can often be controlled externally, which allows the system to be tuned between a slow and a fast regime. In the slow regime, the interdot tunneling rate is slower than the external dephasing rate such that superpositions of charge states are suppressed and the dynamics is well-described by classical models, see, e.g., Refs.~\cite{Kung-PRX-2012,Singh-PRB-2019}. In the fast regime, the tunneling rate is faster than the dephasing rate, permitting the presence of quantum effects \cite{Prech-PRR-2023}. This makes the DQD a highly promising platform for investigating the quantum-to-classical transition.

\begin{figure}
    \centering
    \includegraphics[scale=1]{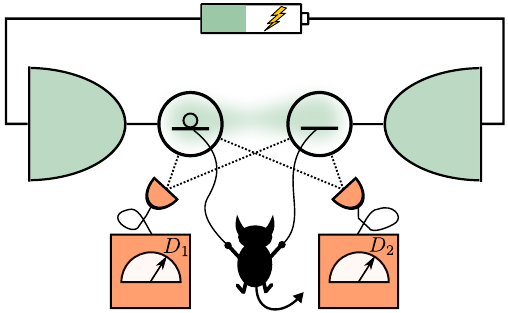}
    \caption{A double quantum dot is continuously monitored by two detectors. A demon uses the outcomes of the measurements to control the on-site energies of the dots, guiding electrons against an applied voltage bias across the terminals. In the ideal cycle, the demon does not perform any net work on the electrons, seemingly violating the second law of thermodynamics.}
    \label{fig:SketchProtocol}
\end{figure}

\begin{figure*}
    \centering
    \includegraphics[scale=0.5]{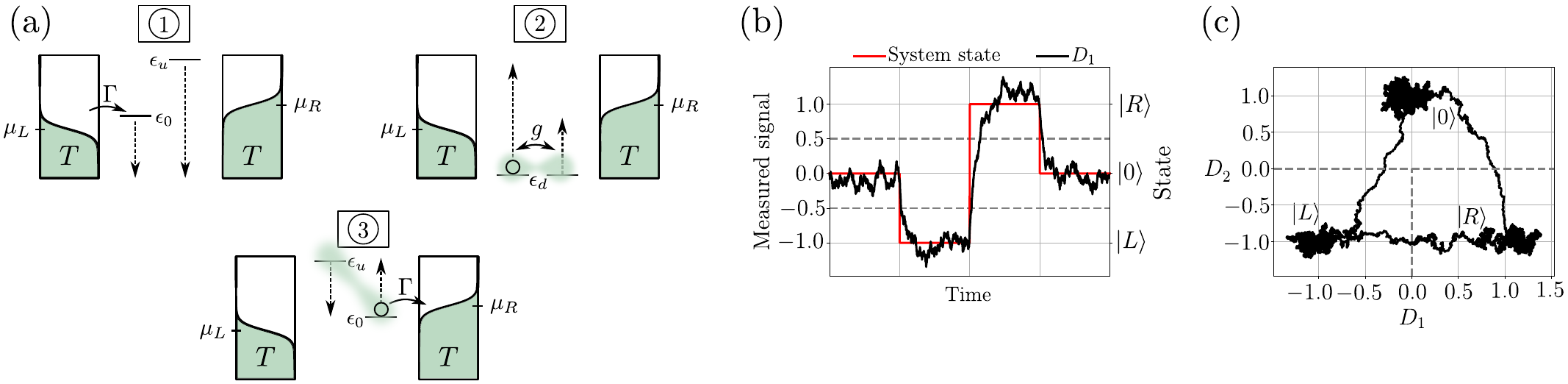}
    \caption{ (a) System and working principle of the demon. The double quantum dot is coupled to electron reservoirs of temperature $T$ and chemical potentials $\mu_{L/R}$. The electron occupation is continuously measured. Solid arrows indicate the desired tunneling events, where $\Gamma$ and $g$ are the tunnel couplings. Dashed arrows visualize the intended feedback actions based on the measurement. (b) Sample trajectory of the system state and the detector signal when using a single detector ($\hat{A}_1=\dyad{R}-\dyad{L}$). The dashed lines indicate the detection strategy, i.e., $\ket{L}$ is assumed to be occupied for $D_1<-1/2$, $\ket{0}$ is assumed to be occupied for $-1/2<D_1<1/2$, and $\ket{R}$ is assumed to be occupied for $D_1>1/2$. Note that the transition $\ket{L}\to\ket{R}$ is interpreted erroneously due to the finite delay time in $D_1$. (c) Sample trajectory when using two detectors [$\hat{A}_1$ and $\hat{A}_2=\dyad{0}-(\dyad{L}+\dyad{R})$] for the same system trajectory as in (b). Dashed lines indicate the detection strategy, where $\ket{0}$ is assumed to be occupied if $(D_1,D_2)$ resides in the two upper quadrants, while $\ket{L}$ is assumed to be occupied when $(D_1,D_2)$ is in the lower left quadrant, and $\ket{R}$ is assumed to be occupied when $(D_1,D_2)$ is located in the lower right quadrant. No erroneous interpretations are observed.}
    \label{fig:System}
\end{figure*}

To date, Maxwell's demon has been experimentally demonstrated both in the classical \cite{Serreli-Nature-2007,Toyabe-Nat.Phys-2010,Koski-PNAS-2014,Chida-Nat.Com.-2017,Kumar-Nature-2018,Ribezzi-Nat.Phys.-2019} as well as the quantum regime \cite{Vidrighin-PRL-2016,Cottet-PNAS-2017,Masuyama-Nat.Com.-2018,Naghiloo-PRL-2018}. However, the quantum-to-classical transition has received much less attention. Here we fill this gap by investigating Maxwell's demon across the quantum-to-classical transition in a DQD, providing insight into the role of quantum coherence in a paradigmatic example of a feedback-controlled system. For an investigation into the quantum-to-classical transition in an autonomous information ratchet (i.e., without feedback), see Ref.~\cite{stevens:2019}.
%In this regard, the DQD is a promising candidate. Additionally, there has not yet been any experimental demonstration of the demon in DQDs.

Averin et.~al. \cite{Averin-PRB-2011} introduced a protocol for implementing Maxwell's demon in a DQD (or metallic islands coupled in series). The protocol is based on a continuous measurement of the charge state and uses this information to guide electrons against an applied bias by manipulating the on-site energies of the dots, see Fig.~\ref{fig:SketchProtocol}. When executed ideally, the process does not change the energy of the electrons, and thus converts heat from the environment into work, only by using information. An additional study of the protocol involved a full statistical characterization of electron transport, as well as a realistic detector model, investigating how detector delay and noise influences the feedback-controlled dynamics \cite{Annby-PRB-2020}. However, these studies concentrated on classical descriptions, neglecting quantum effects altogether.

% Paragraph 5: Summarize how our main results add to our knowledge
In this paper, we extend the operation of this demon to the quantum domain. %In particular, we investigate the quantum-to-classical transition, and highlight the role of quantum effects in information-to-work conversion. 
To model the feedback protocol, we use the quantum Fokker-Planck master equation (QFPME) derived in Ref.~\cite{Annby-PRL-2022}. This equation can model any quantum system under continuous, Markovian feedback. %This equation allows us to study how the strength of the measurement and the bandwidth of the detector influence the dynamics of a feedback-controlled system. 
Under strong dephasing, quantum effects can be neglected and the QFPME reduces to a classical rate equation. Changing the dephasing, for instance by changing the measurement strength, thus allows for probing the quantum-to-classical transition. In the strong measurement regime, we observe the Zeno effect \cite{Misra-JMP-1977,mensky-book-measurements}, where interdot transitions are completely suppressed, quenching the information-to-work conversion.  For weak measurements, we observe a Zeno-like effect, where measurement errors induce dephasing by randomly changing the on-site energies. Compared to a classical device, the measurement strength thus plays a much more intricate role here. As a result, there is no universal quantum advantage or disadvantage. Instead, whether or not quantum coherence is beneficial depends on the involved timescales.

The rest of this paper is structured as follows. Section \ref{sec:WorkingPrinciple} introduces the working principle of the demon. First, we review the classical operation, as introduced in Refs.~\cite{Averin-PRB-2011,Annby-PRB-2020}. Then, we discuss how the operation can be extended to the quantum domain. In Sec.~\ref{sec:ModelEnergetics}, we introduce the QFPME, we show how the energy flows are computed, and we discuss the assumptions that enter our description. In Sec.~\ref{sec:performanc}, we highlight the performance of the demon with qualitative as well as quantitative results. Section \ref{sec:quantum_and_classical} introduces a classical model and identifies when the demon operates in the quantum or in the classical regime. In Sec.~\ref{sec:conclusions} we conclude our discussions and provide possible future research directions.

\section{Working principle}
\label{sec:WorkingPrinciple}

In this section,  we introduce the working principle of the demon. We begin by reviewing the classical description analyzed in Ref.~\cite{Annby-PRB-2020}. When the basic working principle is established, we describe how the system can be pushed into a regime with quantum effects. We also discuss how measurement and feedback can be implemented in a double quantum dot (DQD).

\subsection{Review of classical operation}
\label{sec:ReviewClassical}

Figure \ref{fig:System}\,(a) illustrates the system and working principle. The system consists of a spinless double quantum dot with large intra- and interdot Coulomb repulsion, such that at most one electron resides in the system at all times. Therefore, only three charge states are available; empty, left dot occupied, and right dot occupied. In this section, we assume that the coherence time of the DQD is much shorter than the timescale of its dynamics, such that coherent superpositions between the charge states can be neglected. The dynamics of the system are thus effectively classical. With the aid of voltage gates, the energies of the left and right dot $(\epsilon_L,\epsilon_R)$ can be controlled on a timescale that is much faster than the system dynamics \cite{Barker-PRL-2022}. We focus on the following configurations; $(\epsilon_0,\epsilon_u)$, $(\epsilon_d,\epsilon_d)$, and $(\epsilon_u,\epsilon_0)$, as illustrated in Fig.~\ref{fig:System}\,(a). The system is coupled to two electron reservoirs, each with temperature $T$, but different chemical potentials due to an external voltage bias $\mu_R-\mu_L=eV\geq0$. Throughout the paper, we put $\mu_R=-\mu_L=eV/2$ and $\epsilon_0=0$.

The demon is implemented via a feedback protocol based on continuous measurements of the charge state. To understand the basic working principle, it is instructive to assume that the measurement is ideal such that the charge state is known at all times. The aim of the protocol is to transport electrons against the applied bias, without performing any net work by the voltage gates. This can be achieved in the following way: Initially [(1) in Fig.~\ref{fig:System}\,(a)], the DQD is empty in configuration $(\epsilon_0,\epsilon_u)$. By letting $\epsilon_u-\mu_{L/R}\gg k_BT$, with $k_B$ being the Boltzmann constant, electrons cannot tunnel into the right dot from the right reservoir. When measuring that an electron has tunnelled into the left dot, the level configuration is immediately adjusted to $(\epsilon_d,\epsilon_d)$, as illustrated under (2) in Fig.~\ref{fig:System}\,(a). By choosing $\mu_{L/R}-\epsilon_d \gg k_BT$, the electron remains in the DQD in this configuration. When we detect that the electron tunnels to the right dot, the levels are adjusted to $(\epsilon_u,\epsilon_0)$, as depicted in (3). As the electron tunnels into the right reservoir, we return to the initial configuration, completing one cycle.

For one cycle, the first law of thermodynamics for the DQD reads $W+Q+W_g=0$. Here $W=\mu_L-\mu_R=-eV$ is the chemical work for transporting one electron against the bias. We use the sign convention $W<0$ when the energy of the DQD is decreased, such that the extracted work $-W=eV>0$. The heat exchanged with the reservoirs is given by $Q_L=\epsilon_0-\mu_L$ and $Q_R=\mu_R-\epsilon_0$, giving the total heat $Q=Q_L+Q_R=-W$. Heat is thus converted into work, and with $\mu_L < \epsilon_0 < \mu_R$, both reservoirs are being cooled. Note that the work done by the voltage gates $W_g=0$ since the work extracted when lowering the electron on the left dot is put back into the system when raising the electron on the right dot. Thus, for one cycle, the change of entropy (in system and reservoirs) $\Delta S = -Q/T<0$, seemingly violating the second law of thermodynamics. By including information processing in the thermodynamic book-keeping, a violation is not observed, see for instance Refs.~\cite{Maruyama-RMP-2009} and \cite{Parrondo-Nat.Phys-2015}.

By relaxing the assumption of ideal measurements, i.e., by introducing delay and noise in the charge detection, the performance of the control protocol is modified \cite{Annby-PRB-2020}. Delay allows for electron back-tunneling before applying feedback. This introduces a lag in the operation cycle, reducing the extraction of work. Noise introduces feedback mistakes, which result in heating (and $W_g>0$). When the mistakes become pronounced, the device is operated as an electron pump rather than a demon.

\subsection{Quantum operation}
\label{sec:QuantumOperation}

We now extend the operation into a regime where quantum effects are significant. This can be achieved by increasing the interdot tunnel coupling, such that the system dynamics becomes comparable to the coherence time of the DQD. This introduces new features in the dynamics since coherent superpositions between the left and right charge states can no longer be neglected. We further note that measurement backaction, just like environmental noise, results in dephasing and pushes the operation back toward the classical regime. To minimize backaction, the detector must only weakly interact with the system. Too strong interaction results in a continuous projective measurement, giving rise to the Zeno effect and a complete suppression of interdot tunneling. %From the analysis of Ref.~\cite{Annby-PRB-2020}, we note that noise and delay in the detection will influence the success of the feedback protocol.

To describe the dynamics of the quantum demon, we denote by $\ket{0}$, $\ket{L}$, and $\ket{R}$, the three charge states of the DQD, corresponding to empty, left dot occupied, and right dot occupied, respectively. The Hamiltonian of the system reads
\begin{equation}
\hat{H}(\epsilon_L,\epsilon_R) = \epsilon_L\dyad{L} + \epsilon_R\dyad{R} + g\big(\dyad{L}{R}+\rm h.c.\big),
\label{eq:DQDHamiltonian}
\end{equation}
where $g$ is the interdot tunnel coupling. The DQD is incoherently coupled to electron reservoirs via the transition rate $\Gamma$, as depicted in Fig.~\ref{fig:System}\,(a). The reservoirs are described by the Fermi-Dirac distribution
\begin{equation}
    f_{L(R)}(\epsilon) = [e^{(\epsilon-\mu_{L(R)})/k_BT}+1]^{-1}.
\label{eq:FermiDirac}
\end{equation}
For the remainder of the paper, we set $k_B=1$. In the absence of measurement and feedback, the DQD dynamics are determined by the local Lindblad master equation \cite{Potts-NJP-2021} $\partial_t\hat{\rho}_t=\mathcal{L}(\epsilon_L,\epsilon_R)\hat{\rho}_t$, where
\begin{equation}
\begin{aligned}            &\mathcal{L}(\epsilon_L,\epsilon_R)\hat{\rho} =  -i[\hat{H}(\epsilon_L,\epsilon_R),\hat{\rho}] \\ &+ \sum_{\alpha=L,R} \Big[\gamma_\alpha(\epsilon_\alpha)\mathcal{D}[\dyad{\alpha}{0}]\hat{\rho} + \kappa_\alpha(\epsilon_\alpha)\mathcal{D}[\dyad{0}{\alpha}]\hat{\rho} \Big],
\end{aligned}
\label{eq:DQDDynamics}
\end{equation}
with $\gamma_\alpha(\epsilon)=\Gamma_\alpha f_\alpha(\epsilon)$ and $\kappa_\alpha(\epsilon)=\Gamma_\alpha[1-f_\alpha(\epsilon)]$ being the electron transition rates for jumping in and out of the DQD via bath $\alpha$ at energy $\epsilon$, where $\Gamma_\alpha$ is the bare tunneling rate between the system and bath $\alpha$ (for simplicity assumed equal in the following, $\Gamma=\Gamma_L=\Gamma_R$), and $\mathcal{D}[\hat{c}]\hat{\rho}=\hat{c}\hat{\rho}\hat{c}^\dagger-\frac{1}{2}\{\hat{c}^\dagger\hat{c},\hat{\rho}\}$ for an arbitrary operator $\hat{c}$. For a justification of the local approach used here, see Appendix \ref{appendix:LocalApproach}. Note that we use $\hbar=1$.

The charge state of a DQD can be measured via a probe system that is capacitively coupled to the DQD \cite{Fujisawa-Science-2006,Gustavsson-PRL-2006, Utsumi-PRB-2010,Kung-PRX-2012}. Typically, the probe is a quantum point contact or a single electron transistor. If the coupling is asymmetric with respect to the left and right dot, the electrical current through the probe depends on the charge state of the DQD, and can be used to resolve single electron transitions in real-time. Figure \ref{fig:System}\,(b) illustrates a typical trajectory of such a measurement, where the detector couples to the observable $\hat{A}_1=\dyad{R}-\dyad{L}$. As illustrated in Fig.~\ref{fig:System}\,(b), this measurement may result in ambiguous interpretations of charge transitions due to a finite lag in the detector \cite{Annby-PRB-2020,DanielHolstThesis}. Here we resolve this issue by making simultaneous measurement of $\hat{A}_1$ and $\hat{A}_2=\dyad{0}-\big(\dyad{L}+\dyad{R}\big)$. The second observable corresponds to measuring the total charge of the DQD. Note that $[\hat{A}_1,\hat{A}_2]=0$, allowing the uncertainty of both observables to be small simultaneously. In Fig.~\ref{fig:System}\,(c), we visualize a schematic trajectory of the outcomes $\boldsymbol{D}=(D_1,D_2)$ for the system trajectory in Fig.~\ref{fig:System}\,(b), where $D_2$ is the outcome of detector 2. The following detection and feedback strategy is used; if $D_2>0$, $\ket{0}$ is assumed to be occupied and the energy levels are tuned to configuration 1 [see Fig.~\ref{fig:System}\,(a)]. If $D_2<0$ and $D_1<0$, $\ket{L}$ is assumed to be occupied and level configuration 2 is applied. If $D_2<0$ and $D_1>0$,  $\ket{R}$ is assumed to be occupied, and the DQD is tuned to level configuration 3. The dashed lines in Fig.~\ref{fig:System}\,(c) highlight the borders of these regions. Note that ambiguous interpretations of charge transitions are avoided. In an experiment, the detectors might couple differently to the different charge states, see Appendix \ref{appendix:PostProcess} for a discussion on how our model can be modified in this case.

\section{Model and energetics}
\label{sec:ModelEnergetics}

\subsection{Quantum Fokker-Planck master equation}
\label{sec:QFPME}

To model the feedback protocol, we employ the quantum Fokker-Planck master equation (QFPME) developed in Ref.~\cite{Annby-PRL-2022}, see also Refs.~\cite{Wiseman/Milburn-book-2010,Diosi-PhysScripta-2014,Diosi-PRA-2023,Diotallevi-NJP-2024}. This equation describes the feedback-controlled dynamics of a quantum system undergoing continuous monitoring with finite bandwidth. For the measurement of the DQD, the QFPME can be extended to multiple detectors \cite{DeSousa-InPrep-2024}. It is given by
\begin{equation}
\begin{aligned}
    \partial_t&\hat{\rho}_t(\boldsymbol{D}) = \mathcal{L}(\boldsymbol{D})\hat{\rho}_t(\boldsymbol{D}) + \Tilde{\Gamma}\mathcal{D}[\hat{A}_1]\hat{\rho}_t(\boldsymbol{D}) \\
    &+ \sum_{j=1,2} \left[ \gamma_j \partial_{D_j} \mathcal{A}_j(D_j)\hat{\rho}_t(\boldsymbol{D}) + \frac{\gamma_j^2}{8\lambda_j} \partial_{D_j}^2 \hat{\rho}_t(\boldsymbol{D}) \right],
\end{aligned}
\label{eq:FullQFPME}
\end{equation}
where $\hat{\rho}_t(\boldsymbol{D})$ is the joint state of the DQD and the two detectors. The state of the DQD, independent of the detectors, is given by $\hat{\rho}_t = \int dD_1dD_2\hat{\rho}_t(\boldsymbol{D})$, and $p_t(\boldsymbol{D})=\trace\{\hat{\rho}_t(\boldsymbol{D})\}$ is the joint probability distribution of observing $D_1$ and $D_2$ at time $t$. %The marginal distributions for the individual outcomes are given by $P_t(D_{1(2)}) = \int dD_{2(1)}P_t(\boldsymbol{D})$. 
Note that $\hat{\rho}_t(\boldsymbol{D})$ is normalized according to $\int dD_1dD_2 \trace\{\hat{\rho}_t(\boldsymbol{D})\}=1$. %and that $1=\trace\{\hat{\rho}_t\}$, as well as $1=\int dD_1dD_2 P_t(\boldsymbol{D})$. 

The feedback-controlled dynamics of the DQD discussed in the last section is described by the first term on the rhs of Eq.~(\ref{eq:FullQFPME}), with
\begin{equation}
\begin{aligned}
        \mathcal{L}(\boldsymbol{D}) = \theta(D_2)\mathcal{L}_1 + [1-\theta(D_1)][1-\theta(D_2)]\mathcal{L}_2 \\ + \theta(D_1)[1-\theta(D_2)]\mathcal{L}_3,
\end{aligned}
\label{eq:FullFeedbackLiouvillian}
\end{equation}
where $\theta(x)$ is the Heaviside step function, and $\mathcal{L}_1=\mathcal{L}(\epsilon_0,\epsilon_u)$, $\mathcal{L}_2=\mathcal{L}(\epsilon_d,\epsilon_d)$, and $\mathcal{L}_3=\mathcal{L}(\epsilon_u,\epsilon_0)$ correspond to the dynamics of the three configurations in Fig.~\ref{fig:System}\,(a), with $\mathcal{L}(\epsilon_L,\epsilon_R)$ given by Eq.~(\ref{eq:DQDDynamics}).

The second term of the rhs of Eq.~(\ref{eq:FullQFPME}) represents the dephasing induced by measurements and environmental noise. It describes how the coherence between the states $\ket{L}$ and $\ket{R}$ is damped at a rate given by $\Tilde{\Gamma}=\lambda_1+\Gamma_\varphi$, where $\lambda_1$ is the measurement strength of detector 1, and $\Gamma_\varphi$ is the environmental dephasing rate -- here added phenomenologically~\cite{Kiesslich-PRB-2006}. Since the Hamiltonian (\ref{eq:DQDHamiltonian}) does not create coherence between $\ket{0}$ and $\ket{L/R}$, $[\hat{A}_2,\hat{\rho}_t(\boldsymbol{D})]=0$, implying that there is no backaction associated to the measurement of the total charge.

The terms under the sum in Eq.~(\ref{eq:FullQFPME}) constitute a set of Fokker-Planck equations, and describe the dynamics of the detectors, each with a bandwidth $\gamma_j$. Note that the inverse bandwidth determines the lag of the detector. The superoperator drift coefficient $\mathcal{A}_j(D_j)\hat{\rho}=\frac{1}{2}\{D_j-\hat{A}_j,\hat{\rho}\}$, defines the coupling between the DQD and detector $j$, and determines the average position of the detector, given a certain system state. The diffusion term describes the noise of the detectors, where $\gamma_j/8\lambda_j$ determines the magnitude of the noise. A strong (weak) measurement $\lambda_j\gg\gamma_j$ ($\lambda_j\ll\gamma_j$) reduces (increases) the uncertainty of the measurement. Thus, for detector 1, there is a trade-off between backaction and uncertainty -- a strong measurement minimizes the uncertainty, but destroys the coherence of the DQD. Similarly, a weak measurement preserves coherence, but increases the uncertainty. Such a trade-off is absent for the second detector, as $\hat{A}_2$ is backaction-free.

\subsection{Energetics}
\label{sec:Energetics}

To identify parameter regimes where the feedback protocol can be classified as a Maxwell demon, we evaluate the average stationary energy flows of the DQD. The average energy of the DQD is defined as
\begin{equation}
    E(t) = \int dD_1dD_2 \trace\{\hat{H}(\boldsymbol{D})\hat{\rho}_t(\boldsymbol{D})\},
    \label{eq:AverageEnergyDQD}
\end{equation}
where $\hat{H}(\boldsymbol{D})=\theta(D_2)\hat{H}(\epsilon_0,\epsilon_u)+[1-\theta(D_1)][1-\theta(D_2)]\hat{H}(\epsilon_d,\epsilon_d)+\theta(D_1)[1-\theta(D_2)]\hat{H}(\epsilon_u,\epsilon_0)$ is the feedback Hamiltonian, with $\hat{H}(\epsilon_L,\epsilon_R)$ given in Eq.~(\ref{eq:DQDHamiltonian}). The rate of change of energy is given by $\dot{E}(t)$ [$\dot{E}(t)>0$ when energy enters the DQD], and with the aid of Eq.~(\ref{eq:FullQFPME}), we find, in the stationary limit,
\begin{equation}
    0 = P + \dot{Q} + \dot{E}_{\rm D}.
    \label{eq:1stLawEnsambleLevel}
\end{equation}
Here $P$ and $\dot{Q}$ are the electrical power and heat current associated to transfers of electrons between the reservoirs and the DQD. These energy currents are positive (negative) when energy enters (leaves) the DQD. The power and heat current can be decomposed as (detailed derivation in Appendix \ref{appendix:DetailsEnergetics})
\begin{equation}
    P=\sum_{\alpha=L,R}\sum_{j=1}^3 \mu_\alpha \langle \dot{n} _{\alpha}^{(j)}\rangle, 
    \label{eq:EnsemblePower}
\end{equation}
and
\begin{equation}
    \dot{Q}=\sum_{\alpha=L,R}\sum_{j=1}^3 (\epsilon_\alpha^{(j)}-\mu_\alpha) \langle \dot{n} _{\alpha}^{(j)}\rangle,
    \label{eq:EnsembleHeatCurrent}
\end{equation}
where $\epsilon_\alpha^{(j)}$ is the energy of dot $\alpha$ in level configuration $j$, and
\begin{equation}
\begin{aligned}
    \langle\dot{n}_{\alpha}^{(j)}\rangle = \int &dD_1dD_2 h_j(\boldsymbol{D})\\ &\times[\gamma_\alpha(\epsilon_\alpha^{(j)})\rho_{00}(\boldsymbol{D}) - \kappa_\alpha(\epsilon_\alpha^{(j)})\rho_{\alpha\alpha}(\boldsymbol{D}) ]
\end{aligned}
\label{eq:ParticleCurrents}
\end{equation}
is the average particle current exchanged between the DQD and bath $\alpha$ in level configuration $j$, with $\rho_{\alpha\alpha}(\boldsymbol{D})=\mel{\alpha}{\hat{\rho}_{\rm ss}(\boldsymbol{D})}{\alpha}$, for the stationary density matrix $\hat{\rho}_{\rm ss}(\boldsymbol{D})$, and
\begin{equation}
    h_j(\boldsymbol{D}) =
    \begin{cases}
    \theta(D_2), \hspace{3cm} j=1, \\
    [1-\theta(D_1)][1-\theta(D_2)], \hspace{0.5cm} j=2, \\
    \theta(D_1)[1-\theta(D_2)], \hspace{1.3cm} j=3,
    \end{cases}
    \label{eq:FeedbackRegions}
\end{equation}
is a function that limits the integration for configuration $j$. It is thus possible to infer $P$ and $\dot{Q}$ by counting the number of electrons exchanged between the DQD and the baths in each level configuration.

The current $\Dot{E}_{\rm D}$ corresponds to all energy contributions provided by the demon through gate operations and measurements. It is inferred directly from Eq.~(\ref{eq:1stLawEnsambleLevel}),
\begin{equation}
    \dot{E}_{\rm D} = -P-\dot{Q}.
    \label{eq:EnsembleDemonEnergy}
\end{equation}
In Appendix \ref{appendix:DetailsEnergetics}, we explicitly show that the gate work is the main contribution to $\dot{E}_{\rm D}$. In addition, we detail the derivation of all energy currents, and show how they can be calculated numerically with Eq.~(\ref{eq:QFPME_SepTS_D2}).
To classify the feedback protocol as a demon, we require that $\dot{Q}>0$, such that heat is extracted from the reservoir (seemingly violating the second law) and converted into work. %This implies, together with Eq.~(\ref{eq:1stLawEnsambleLevel}), that $P+\dot{E}_{\rm D} < 0$, ensuring that the major part of the extracted power is provided by heat from the reservoirs, rather than from measurements and gate work. 
We are mainly interested in the regime where $\dot{E}_{\rm D}$ is small, such that heat from the reservoirs is converted into electrical power using information.

We note that $\dot{Q}>0$ implies that entropy has to be produced elsewhere, i.e., in the detection process. Often the actual entropy produced by the measurement is irrelevant, instead it is the acquired information that determines the limits in work extraction and the efficiency in the information-to-work conversion process \cite{Sagawa-Prog.Theor.Phys-2012,Parrondo-Nat.Phys-2015,Goold-J.Phys.A-2016}. We leave the quantification of information in the quantum Fokker-Planck master equation for future work.

\subsection{Simplifying assumptions}
\label{sec:simplifying_assumptions}

To facilitate the analysis of the demon, it is useful to consider three simplifying assumptions. These are discussed and compared in detail below.

\subsubsection{Ideal charge detection}

 The first assumption relies on the fact that $\hat{A}_2$ is backaction free. This allows us to assume that the second detector is noise-free ($\lambda_2/\gamma_2\rightarrow\infty$) and infinitely fast ($\gamma_2\gg\Gamma,\Tilde{\Gamma},g,|\epsilon_{u/d}-\epsilon_0|$), without affecting the dynamics of the DQD. Under these limits, Eq.~(\ref{eq:FullQFPME}) can be reduced to a one-dimensional QFPME amenable for 
 efficient numerical integration (details in Appendix \ref{appendix:FastDetectors}),
\begin{equation}
\begin{aligned}
    \partial_t\hat{\rho}_t(D_1) &= \Tilde{\mathcal{L}}(D_1)\hat{\rho}_t(D_1) + \Tilde{\Gamma}\mathcal{D}[\hat{A}_1]\hat{\rho}_t(D_1) \\
    &+ \gamma_1 \partial_{D_1} \mathcal{A}_1(D_1)\hat{\rho}_t(D_1) + \frac{\gamma_1^2}{8\lambda_1} \partial_{D_1}^2\hat{\rho}_t(D_1),
\end{aligned}
\label{eq:QFPME_SepTS_D2}
\end{equation}
where $\hat{\rho}_t(D_1)=\int dD_2\hat{\rho}_t(\boldsymbol{D})$, and $\Tilde{\mathcal{L}}(D_1)=\mathcal{L}_1\mathcal{V}_{00} + \{ [1-\theta(D_1)]\mathcal{L}_2 + \theta(D_1)\mathcal{L}_3 \}\left(1-\mathcal{V}_{00} \right)$ describes the feedback-controlled dynamics, with the projection superoperator $\mathcal{V}_{aa'}\hat{\rho}\equiv\mel*{a}{\hat{\rho}}{a'}\dyad*{a}{a'}$ for system states $\ket{a}$ and $\ket{a'}$. We note that these assumptions imply that electrons only enter the DQD in configuration 1, and only leave the DQD in configurations 2 and 3, as the energy levels are immediately adjusted when an electron enters or leaves the DQD.
Nevertheless, this model allows us to investigate the main features of the DQD demon across the quantum to classical transition. An investigation into a non-ideal total charge detector is left for future work.
For details related to the numerical integration of Eq.~(\ref{eq:QFPME_SepTS_D2}), we refer to Appendix \ref{appendix:NumericalMethod}. From here on, we focus on ideal charge detection throughout the paper unless otherwise stated.

\subsubsection{Fast detectors}
\label{sec:fast_detectors}

For fast detectors ($\gamma_1,\gamma_2\gg\Gamma,\Tilde{\Gamma},g,|\epsilon_{u/d}-\epsilon_0|$), Eq.~(\ref{eq:QFPME_SepTS_D2}) can be reduced to a Markovian master equation for the DQD, independent of the detectors. This equation reads (derivation in Appendix \ref{appendix:FastDetectors})
\begin{equation}
    \partial_t\hat{\rho}_t = \mathcal{L}_{\rm fb} \hat{\rho}_t + \Tilde{\Gamma}\mathcal{D}[\hat{A}_1]\hat{\rho}_t,
    \label{eq:SepTSMasterEquation}
\end{equation}
where the feedback-controlled dynamics are described by $\mathcal{L}_{\rm fb} = \mathcal{L}_1\mathcal{V}_{00} + [(1-\eta)\mathcal{L}_2+\eta\mathcal{L}_3]\mathcal{V}_{LL}+[\eta\mathcal{L}_2+(1-\eta)\mathcal{L}_3]\mathcal{V}_{RR} + (\mathcal{L}_2+\mathcal{L}_3)(\mathcal{V}_{LR}+\mathcal{V}_{RL})/2$, with the feedback error probability 
\begin{equation}
\eta=\frac{1}{2}\left[1-\text{erf}\left(2\sqrt{\lambda_1/\gamma_1}\right)\right].
\label{eq:FeedbackProbError}
\end{equation}
Note that $0\leq\eta\leq1/2$, where the lower bound is satisfied for a noiseless detector ($\lambda_1\gg\gamma_1$), and the upper bound corresponds to a noisy detector ($\lambda_1\ll\gamma_1$). After integrating over the detector variables, it is no longer possible to resolve the individual level configurations. Instead, the effect of feedback enters Eq.~(\ref{eq:SepTSMasterEquation}) via modifications of the bath couplings and a rescaling of the level detuning by a factor $1/2$, i.e., $\epsilon_u-\epsilon_0\to(\epsilon_u-\epsilon_0)/2$. We refer the reader to Appendix \ref{appendix:FastDetectors} for detailed expressions.

\subsubsection{Energy-conserving demon}
\label{sec:energy_conserving}

Our third simplifying assumption consists of suppressing all remaining undesired tunneling events which might occur at energies $\epsilon_{u/d}$ by setting
\begin{equation}
\gamma_L(\epsilon_{u/d})=\kappa_L(\epsilon_{u/d})=\gamma_R(\epsilon_{u/d})=\kappa_R(\epsilon_{u/d})=0.
    \label{eq:ConditionsEnergyConservingDemon}
\end{equation}
This could, in principle, be obtained by using electron reservoirs where the spectral densities have finite bandwidths and vanish at $\epsilon_{u/d}$. Together with the first assumption, the analysis considerably simplifies and allow us to focus on the trade-off between measurement backaction and information gain. By suppressing these tunneling events, electrons can only enter the DQD at $\epsilon_0$ on the left side and leave the DQD at $\epsilon_0$ on the right side. Therefore, the demon is not performing any net work on the electrons ($\Dot{E}_{\rm D}=0$), thus conserving their energy.

\section{Performance of the demon}
\label{sec:performanc}
\subsection{Qualitative behavior}

\begin{figure*}
    \centering
    \includegraphics[scale=0.9]{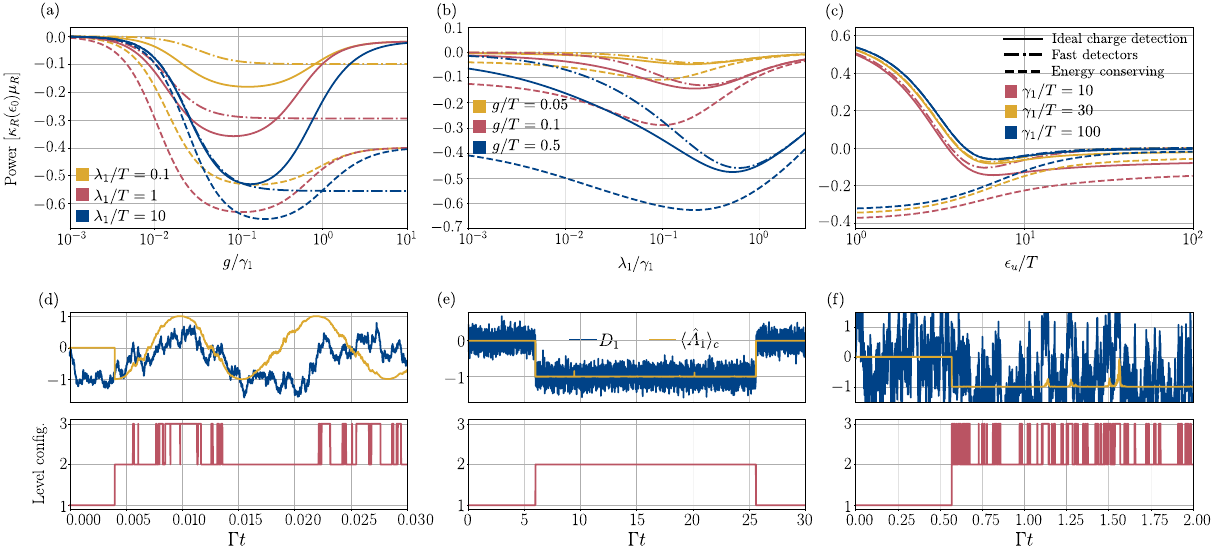}
    \caption{(top) The stationary power (\ref{eq:EnsemblePower}) as a function of (a) $g$, (b) $\lambda_1$, and (c) $\epsilon_u$. The panels compare the simplifying assumptions in Sec.~\ref{sec:simplifying_assumptions}. All lines were calculated by assuming ideal charge detection. Solid lines correspond to Eq.~(\ref{eq:QFPME_SepTS_D2}) (ideal charge detection), dash-dotted lines to Eq.~(\ref{eq:SepTSMasterEquation}) (fast detectors), and dashed lines to Eq.~(\ref{eq:QFPME_SepTS_D2}) using the conditions in Eq.~(\ref{eq:ConditionsEnergyConservingDemon}) (energy-conserving demon). Solid lines are referred to as `Ideal charge detection' as the calculations did not rely on any further assumptions. (a) The power production ($-P$) initially increases with $g$ as more electrons can traverse the sytem per unit time, but decreases when the detector lags behind the interdot tunneling. (b) The power production ($-P$) initially increases with $\lambda_1$ as the noise, and thereby the number of feedback mistakes, reduces. For strong measurements, the power vanishes because of the quantum Zeno effect. (c) For weak measurements, detector noise induces feedback errors, leading to phase damping as the detuning between the left and right dot varies randomly. Parameters: $\Gamma/T=0.1$, $(\mu_R-\mu_L)/T=3$, $\epsilon_0/T=0$, $\Gamma_\varphi/T=0$, and $\epsilon_d=-\epsilon_u$ were used in all plots. For (a) and (b): $\gamma_1/T=10$, $\epsilon_u/T=5$. For (c): $\lambda_1/T=1$, $g/T=0.1$. (bottom) Example trajectories of $D_1$ (blue), the conditional average $\langle\hat{A}_1\rangle_c$ (yellow), and the level configuration (red). For all simulations, we used $Tdt=10^{-4}$. (d) The trajectories highlight how the lag of the detector causes a delay in adjusting the level configuration in the region $0.015<\Gamma t<0.025$. Here we used the same parameters as for the red line in (a) ($\lambda_1/T=1$) with $g/\gamma_1=2.5$. (e) Illustration of the Zeno effect on trajectory level. The strong measurement prevents oscillations between the left and right charge states. We used the same parameters as for the red line ($g/T=0.1$) in (b) with $\lambda_1/\gamma_1=8$. (f) The trajectories illustrate how the levels are randomly shuffled for weak measurements due to the noisy measurement record. The shuffling leads to a Zeno-like situation, where tunneling between the left and right charge states cannot occur. We used the same parameters as in the top panels, with $\lambda_1/T=1$, $g/T=0.1$, and $\epsilon_u/T=5$.} 
    \label{fig:performance_simplifying_assumptions}
\end{figure*}

We begin by discussing the performance of the demon under the simplifying assumptions in Sec.~\ref{sec:simplifying_assumptions}.% To this end, we plot the power (\ref{eq:EnsemblePower}) as a function of (a) $g$, (b) $\lambda_1$, and (c) $\epsilon_u$ in Fig.~\ref{fig:performance_simplifying_assumptions}. Solid lines were calculated with Eq.~(\ref{eq:QFPME_SepTS_D2}). We refer to these lines as the ideal charge detection model in the figure, as no further assumptions were made to obtain them. Therefore, the solid lines act as a benchmark for the dash-dotted and dashed lines, as the latter rely on additional assumptions. The dash-dotted lines were calculated with Eq.~(\ref{eq:SepTSMasterEquation}), and are referred to as the fast-detector model. The dashed lines were calculated with Eq.~(\ref{eq:QFPME_SepTS_D2}) using the conditions in Eq~(\ref{eq:ConditionsEnergyConservingDemon}), and we refer to these lines as the energy conserving model.
To this end, we consider the power produced by the demon as we vary the inter-dot tunneling $g$, the measurement strength $\lambda_1$, as well as the splitting between the dot energies $\epsilon_u$ as illustrated in Fig.~\ref{fig:performance_simplifying_assumptions}(a)-(c).
As $g$ is increased, the power initially grows as more electrons can traverse the system per unit time, see Fig.~\ref{fig:performance_simplifying_assumptions}\,(a). When $g\gtrsim\gamma_1$, the detector starts to lag behind the interdot tunneling, inducing feedback mistakes, and thus reducing the power. %The same behavior is observed for the energy-conserving demon (dashed lines). The discrepancy between solid and dashed lines is due to the conditions in Eq.~(\ref{eq:ConditionsEnergyConservingDemon}), prohibiting tunneling between the baths and the DQD at energies $\epsilon_{u/d}$, thus allowing higher power for the energy-conserving demon. 
The fast-detector model does not capture this non-monotonic behavior, but rather reaches a constant power for $g/\gamma_1\gtrsim10^{-1}$, where the fast-detector assumption breaks down. %For all models, we note that slightly increasing $\lambda_1$ results in higher power. However, as illustrated in Fig.~\ref{fig:performance_simplifying_assumptions}\,(b), increasing $\lambda_1$ indefinitely is detrimental because of the quantum Zeno effect.

Figure \ref{fig:performance_simplifying_assumptions}\,(b) illustrates that the power initially increases with $\lambda_1$ as the noise of detector 1, and thereby also the number of feedback mistakes, reduces. For strong measurements ($\lambda_1\gg\gamma_1$), the power goes to zero due to the quantum Zeno effect, prohibiting interdot tunneling, see also Ref.~\cite{Schaller-PRB-2018}, where the Zeno effect was observed for a similar quantum dot system. We further note that all models merge for strong measurements. In this regime, feedback mistakes are heavily suppressed, such that all electrons that enter the left dot at $\epsilon_0$ leave the right dot at the same energy. Therefore, there is good agreement between the ideal charge detector and energy conserving models. The fast-detector model performs well as strong measurements slow down interdot transitions, making the fast-detector assumption a good approximation.

Figure \ref{fig:performance_simplifying_assumptions}\,(c) highlights that the power vanishes for large $\epsilon_u$ when the detectors are noisy ($\gamma_1\gg\lambda_1$). The noise induces feedback mistakes, and leads to rapid random changes between level configuration 2 and 3 when an electron resides in the DQD. The detuning between the left and right level is thus randomly changing, giving rise to phase damping \cite{nielsen-chuang-book} and the power vanishes. For smaller $\gamma_1$, the phase damping is less pronounced as the measurement becomes less noisy.

\begin{figure*}
    \centering
    \includegraphics[width=.9\textwidth]{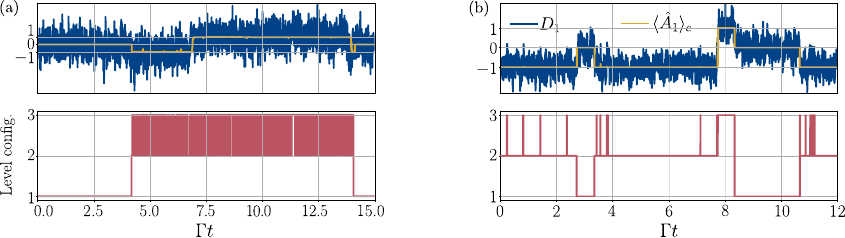}
    \caption{Example trajectories for different measurement strengths: (a) $\lambda_1/\gamma_1 = 0.1$, (b) $\lambda_1/\gamma_1 = 1$. Other parameters are as in Fig.~\ref{fig:performance_simplifying_assumptions}. A weak measurement results in many errors in the feedback, see panel (a). An intermediate value for $\lambda_1$ limits the number of errors while still avoiding the Zeno effect illustrated in Fig.~\ref{fig:performance_simplifying_assumptions}(e).}
    \label{fig:trajslambda}
\end{figure*}

Figures \ref{fig:performance_simplifying_assumptions}(d)-(f) show individual quantum trajectories $\zeta_\tau=\{\hat{\rho}_c(t),t\in[0,\tau]\}$ of the system, where the density matrix $\hat{\rho}_c(t)$ is conditioned on the full trajectories of measurement outcomes and electron exchanges with the baths up to time $t$. The joint system-detector state of Eq.~(\ref{eq:FullQFPME}) is related to he conditional state via $\hat{\rho}_t(\boldsymbol{D})=\text{E}[\delta\{D_1-D_1(t)\}\delta\{D_2-D_2(t)\}\hat{\rho}_c(t)]$, where $\text{E}[\cdot]$ is the ensemble average over all possible measurement outcomes and electron exchanges with the baths, and the delta functions single out the outcomes at time $t$. Appendix \ref{appendix:MCWF} details the calculations of $\hat{\rho}_c(t)$. We plot the outcome of detector 1 $D_1$, the conditional average $\langle\hat{A}_1\rangle_c=\trace\{\hat{A}_1\hat{\rho}_c(t)\}$, and the level configuration. Figure \ref{fig:performance_simplifying_assumptions}(d) highlights how the lag of the detector causes a delay in adjusting the level configuration. Such delays lead to a reduction in power as illustrated in Fig.~\ref{fig:performance_simplifying_assumptions}(a). Figure \ref{fig:performance_simplifying_assumptions}(e) illustrates the Zeno effect. Strong measurements suppress tunneling between the left and right charge states, prohibiting power production. Figure \ref{fig:trajslambda} further illustrates the effect of the measurement strength on the trajectories. An intermediate value of $\lambda_1$ limits the number of mistakes in the feedback while also preventing the Zeno effect. In Fig.~\ref{fig:performance_simplifying_assumptions}(f), we highlight how the levels are randomly shifted between configuration 2 and 3 for weak measurements, as the trajectory of $D_1$ becomes noisy. Similar to the Zeno effect, this prevents tunneling between the left and right charge states, suppressing the power.

\subsection{Quantitative behavior for fast detectors}
For fast detectors, it is possible to write down analytical expressions for the power (\ref{eq:EnsemblePower}) and the heat current (\ref{eq:EnsembleHeatCurrent}). The derivation relies on the assumption $|\epsilon_{u/d}-\mu_{L/R}|\gg T$, which ensures that $\gamma_{L/R}(\epsilon_u),\kappa_{L/R}(\epsilon_d)\ll1$. Under this assumption, we find (see details in Appendix \ref{appendix:FastDetectors})
\begin{equation}
    P = \frac{(\mu_L-\mu_R)(1-\eta)\gamma_L\xi\kappa_R}{[\gamma_L+\eta\kappa_L][\xi+(1-\eta)\kappa_R]+\xi[\gamma_L+(1-\eta)\kappa_R]},
    \label{eq:sepTSPower}
\end{equation}
and
\begin{equation}
    \dot{Q} = -\frac{(\epsilon_u-\epsilon_0)\eta\kappa_L\gamma_L}{\gamma_L+\eta\kappa_L+\xi\frac{\gamma_L+(1-\eta)\kappa_R}{\xi+(1-\eta)\kappa_R}} - P,
    \label{eq:sepTSHeatCurrent}
\end{equation}
with $\gamma_L=\gamma_L(\epsilon_0)$, $\kappa_L=\kappa_L(\epsilon_u)$, $\kappa_R=\kappa_R(\epsilon_0)$, and where we introduced the effective interdot tunneling rate 
\begin{equation}
    \xi = \frac{8g^2(\kappa_L+\kappa_R+8\Tilde{\Gamma})}{(\kappa_L+\kappa_R+8\Tilde{\Gamma})^2+4(\epsilon_u-\epsilon_0)^2}.
    \label{eq:sepTSInterdotRate}
\end{equation}
Despite being valid only for fast detectors, Eqs.~(\ref{eq:sepTSPower}) and (\ref{eq:sepTSHeatCurrent}) provide surprisingly general insights of the feedback protocol. The first term in $\dot{Q}$ corresponds to the heating of the reservoirs when the demon makes feedback mistakes. That is, electrons that enter the left dot ($\gamma_L$) in configuration 1 can dissipate ($\eta\kappa_L$) into the left reservoir in configuration 3, heating the left reservoir. The first term mainly plays a role for weak measurements ($\lambda_1\ll\gamma_1$), where $\eta$ is finite. For strong measurements ($\lambda_1\gg\gamma_1$), the heating effect is quenched. Note that $\eta$ goes to zero exponentially, while $\xi$ goes to zero as $1/\lambda_1$ [cf.~Eqs.~(\ref{eq:FeedbackProbError}) and (\ref{eq:sepTSInterdotRate})], such that $\dot{Q}\simeq -P$, demonstrating perfect heat-to-work conversion when feedback is applied correctly at all times. In the extreme limit $\lambda_1\to\infty$, $P,\dot{Q}\to 0$ due to the quantum Zeno effect, illustrating that interdot tunneling is prohibited ($\xi\to0$). The heating effect for weak measurements can also be quenched by putting $\kappa_L(\epsilon_u)=0$ (satisfying one of the conditions for the energy-conserving demon), eliminating the first term of $\dot{Q}$. By imposing this condition, $\dot{Q}=-P$, highlighting that we can achieve perfect heat-to-work conversion even when $\lambda_1\to0$. That is, the information gained by the ideal total charge measurement is enough to operate the system as Maxwell's demon, even when randomly switching between configuration 2 and 3. We also note that the interdot rate (\ref{eq:sepTSInterdotRate}) is suppressed for large detuning $\epsilon_u-\epsilon_0$, capturing the phase damping effect discussed above.

The condition $\dot{Q}>0$, together with Eq.~(\ref{eq:sepTSHeatCurrent}), implies that the feedback protocol corresponds to a Maxwell demon when
\begin{equation}
    \frac{\mu_R-\mu_L}{\epsilon_u-\epsilon_0} > \frac{\eta}{1-\eta}\frac{\kappa_L[\xi+(1-\eta)\kappa_R]}{\xi\kappa_R}.
    \label{eq:SepTSDemonInequality}
\end{equation}
A large separation between detuning and bias ($\epsilon_u-\epsilon_0\gg\mu_R-\mu_L$) requires strong measurements to satisfy the inequality ($\eta\to0$ for $\lambda_1\to\infty$). By increasing the detuning indefinitely, $\xi\to0$, prohibiting interdot tunneling. The rhs will thus diverge, and violate Eq.~(\ref{eq:SepTSDemonInequality}). When reducing $\lambda_1$, i.e., increasing $\eta$, the inequality is ensured by chosing a smaller detuning. The limit $\epsilon_u-\epsilon_0\ll\mu_R-\mu_L$ appears useful as the lhs becomes large. However, with increased bias, $\kappa_R$ goes to zero while $\kappa_L$ and $\xi$ remain finite, leading to a violation of Eq.~(\ref{eq:SepTSDemonInequality}), as the rhs diverges. We thus understand that the externally controlled parameters $\epsilon_{u/0/d}$ and $\mu_{L/R}$ cannot be chosen arbitrarily, as the rhs diverges for both large detuning and large bias.

\begin{figure*}
    \centering
    \includegraphics[scale=0.95]{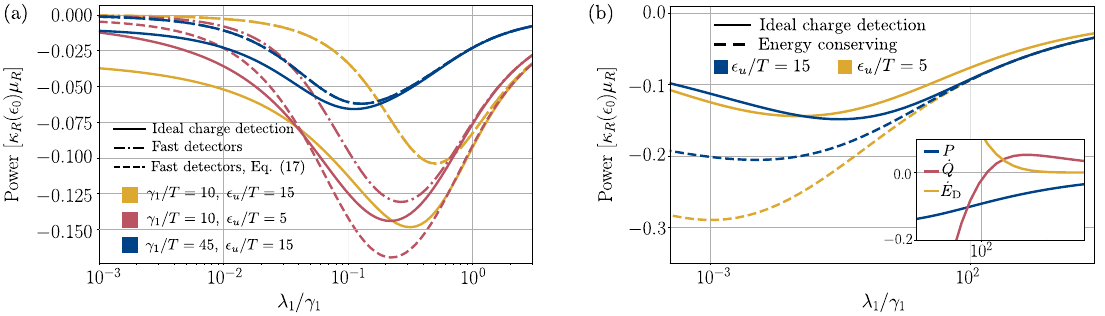}
    \caption{(a) Power as function of $\lambda_1$ for various choices of $\gamma_1$ and $\epsilon_u$. The plot explores the validity of the fast-detector assumption. Solid and dash-dotted lines were calculated numerically with Eqs.~(\ref{eq:QFPME_SepTS_D2}) and (\ref{eq:SepTSMasterEquation}), respectively, while dashed lines correspond to Eq.~(\ref{eq:sepTSPower}). The figure illustrates how Eq.~(\ref{eq:SepTSMasterEquation}) breaks down  when $\gamma_1$ becomes comparable to, or smaller than, the largest inverted timescale of the system. Equation (\ref{eq:sepTSPower}) is valid as long as $|\epsilon_{u/d}-\mu_{L/R}|\gg T$. For $\epsilon_u/T=15$, the dash-dotted and dashed lines completely overlap. Parameters: $\Gamma/T=0.1$, $g/T=0.1$, $(\mu_R-\mu_L)/T=3$, $\epsilon_0/T=0$, $\Gamma_\varphi/T=0$, and $\epsilon_d=-\epsilon_u$. (b) Power (\ref{eq:EnsemblePower}) as a function of $\lambda_1$. The plot compares the ideal charge detector model (\ref{eq:QFPME_SepTS_D2}) and the energy conserving model. The two models coincide for strong measurements when $\epsilon_u$ (and $\epsilon_d)$ are much larger than the temperature of the baths. Under these conditions, the number of feedback mistakes is suppressed and many of the conditions in Eq.~(\ref{eq:ConditionsEnergyConservingDemon}) are approximately fulfilled. Inset: power, heat current and demon energy current as a function of $\lambda_1$ for $\epsilon_u/T=15$. For strong measurements, $\Dot{E}_{\rm D}\simeq0$ and $P\simeq-\dot{Q}$, such that no external work is applied on the electrons. Parameters: same as (a) with $\gamma_1/T=10$.}
    \label{fig:BreakdownSepTS}
\end{figure*}

\subsection{Validity of the approximations}

In Fig.~\ref{fig:BreakdownSepTS}(a), we illustrate the fast-detector assumption, by plotting the power (\ref{eq:EnsemblePower}) as a function of $\lambda_1$. %The solid lines were calculated numerically with Eq.~(\ref{eq:QFPME_SepTS_D2}) and act as a benchmark for the remaining lines. The dash-dotted lines were calculated numerically with Eq.~(\ref{eq:SepTSMasterEquation}). The dashed lines correspond to Eq.~(\ref{eq:sepTSPower}). 
The figure highlights that the fast-detector model is valid as long as $\gamma_1$ is much larger than the remaining inverted timescales -- note that the discrepancy between the models reduces when $\gamma_1$ increases. Additionally, we note that Eq.~(\ref{eq:sepTSPower}) is valid as long as $|\epsilon_{u/d}-\mu_{L/R}|\gg T$ is satisfied -- for $\epsilon_u/T=15$, the full fast-detector model (\ref{eq:SepTSMasterEquation}) and Eq.~(\ref{eq:sepTSPower}) completely overlap.

Figure \ref{fig:BreakdownSepTS}(b) investigates when the energy-conserving demon is a good approximation. To this end, we plot the power (\ref{eq:EnsemblePower}) as a function of $\lambda_1$ by solving Eq.~(\ref{eq:QFPME_SepTS_D2}) with (solid lines) and without (dashed lines) the conditions in Eq.~(\ref{eq:ConditionsEnergyConservingDemon}). We find that there is a good agreement between the models for strong measurements when $\epsilon_u/T=15$. Choosing $\epsilon_u/T=15$ ensures that several of the conditions in Eq.~(\ref{eq:ConditionsEnergyConservingDemon}) are met, prohibiting electrons tunneling into the DQD at $\epsilon_u$ in configuration 1, and out of the DQD at $\epsilon_d$ in configuration 2. Strong measurements ensure that feedback mistakes are suppressed, such that all electrons entering the left dot at $\epsilon_0$ are transported to the right reservoir at $\epsilon_0$. Therefore, the energy of the electrons are conserved, explaining the agreement of the models. For $\epsilon_u/T=5$, none of the conditions in Eq.~(\ref{eq:ConditionsEnergyConservingDemon}) are met, allowing electrons to tunnel into the DQD at $\epsilon_u$ in configuration 1, and into the reservoirs at $\epsilon_d$ in configuration 2. Therefore, the ideal charge detector model predicts a lower power than the energy conserving model. The inset of Fig.~\ref{fig:BreakdownSepTS}(b) highlights how the heat current and demon energy current behave for the ideal charge detection model. For the inset we choose $\epsilon_u/T=15$. For strong measurements, we note that $\dot{E}_{\rm D}\simeq0$, such that $P\simeq -\dot{Q}$. That is, by suppressing feedback mistakes, the protocol in Fig.~\ref{fig:System}\,(a) can be followed perfectly, and the demon does not provide any energy to the DQD. For weak measurements, this is not the case. In particular, we note that $\dot{E}_{\rm D}>0$ while $\dot{Q}<0$. That is, for weak measurements, the demon controls the DQD erroneously, heating the left reservoir, as discussed above. As $\lambda_1\to0$, $\dot{Q}\simeq -\dot{E}_{\rm D}$. Note that neither $\dot{Q}$ nor $\dot{E}_{\rm D}$ diverge in this limit, see Eq.~(\ref{eq:sepTSHeatCurrent}). Note that for the energy-conserving demon, $\dot{E}_{\rm D}=0$ and $P=-\dot{Q}$ always.

\section{Quantum-to-classical transition}
\label{sec:quantum_and_classical}

Under strong dephasing -- here caused by environmental noise and continuous measurements -- it is appropriate to expect that the interdot tunneling is well-modeled by classical jumps, see, e.g., Refs.~\cite{Kung-PRX-2012,Singh-PRB-2019}. To this end, we use our microscopic model (\ref{eq:QFPME_SepTS_D2}) to derive a classical equation of motion of the system. We identify the parameter regimes in which the system is expected to be well-described by the classical equation and we highlight deviations from the classical model as we cross the quantum-to-classical transition into a regime where quantum coherence cannot be neglected. Additionally, we find that the phenomenological description of the demon used in Ref.~\cite{Annby-PRB-2020} emerges from the classical model. 

\begin{figure*}
    \centering
    \includegraphics[scale=1]{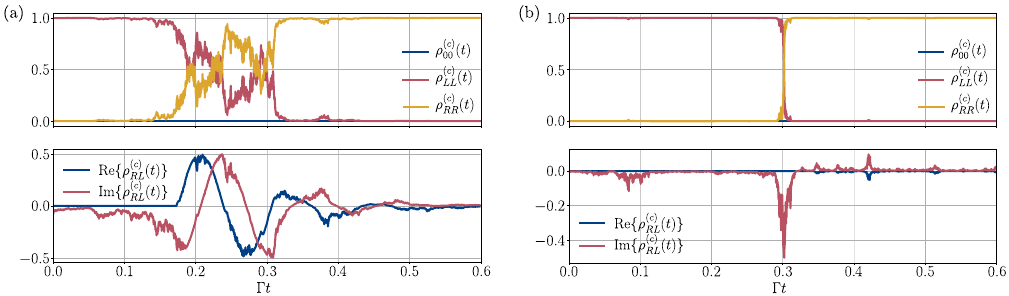}
    \caption{Individual quantum trajectories of $\hat{\rho}_c(t)$ for the parameters in Fig.~\ref{fig:QuantumToClassicalPanel}(b), using $\lambda_1/\gamma_1=0.6$ in (a) and $\lambda_1/\gamma_1=6$ in (b), for both panels, we used $Tdt=10^{-4}$. Here $\rho_{ab}^{(c)}(t)=\mel{a}{\hat{\rho}_c(t)}{b}$. For weak measurements (a), interdot tunneling is accompanied by oscillations in the off-diagonal elements. For strong measurements (b), interdot tunneling resembles a classical jump with a short outburst in the coherence.}
    \label{fig:QtoCTrajectories}
\end{figure*}

We begin by motivating when interdot tunneling is well-described by classical jumps. In Fig.~\ref{fig:QtoCTrajectories}, we plot individual quantum trajectories of the system. Appendix \ref{appendix:MCWF} details the calculations of the trajectories. Figures \ref{fig:QtoCTrajectories}(a) and (b) illustrates interdot transitions when the DQD is operated as a Maxwell demon for weak and strong measurements. For weak measurements (a), interdot transitions are accompanied by oscillations in the off-diagonal density matrix elements, as expected for coherent tunneling. When increasing the measurement strength (b), coherence is heavily suppressed. In particular, we note that the transition becomes sharper in time with a short outburst of coherence. This resembles a classical jump, and motivates why tunneling may be well-modeled by classical jumps.

\begin{figure*}[t!]
\centering
    \includegraphics[scale=1]{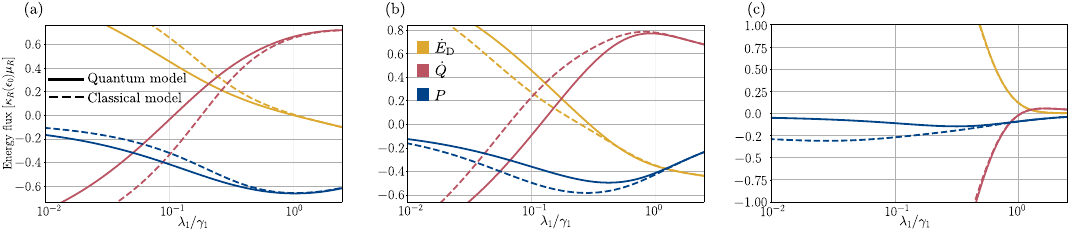}
    \caption{Comparison of the quantum (solid lines) and classical (dashed lines) models. All lines were calculated numerically (App.~\ref{appendix:NumericalMethod}) using Eqs.~(\ref{eq:QFPME_SepTS_D2}) and (\ref{eq:ClassicalQFPME}) for the quantum and classical models, respectively. We plot $P$ (\ref{eq:EnsemblePower}), $\dot{Q}$ (\ref{eq:EnsembleHeatCurrent}), and $\dot{E}_{\rm D}$ (\ref{eq:EnsembleDemonEnergy}) as functions of $\lambda_1$. For a slow detector, (a) $\gamma_1/T=0.1$, the quantum model predicts more extracted power while for a fast detector, (b) $\gamma_1/T=1$ and (c) $\gamma_1/T=10$, the classical model predicts a higher extracted power due to the feedback-induced phase damping effect, see main text. The models merge for strong measurements, where measurement backaction suppresses quantum coherence. Panel (c) ($\epsilon_u/T=15$) illustrates that the demon energy only vanishes for strong measurements, where the classical model captures the dynamics. Parameters (unless otherwise stated): $\Gamma/T=0.1$, $g/T=0.1$, $\gamma_1/T=0.1$, $\epsilon_u/T=5$, $\epsilon_0/T=0$, $(\mu_R-\mu_L)/T=3$, $\Gamma_\varphi/T=0$, $\epsilon_d=-\epsilon_u$.}
    \label{fig:QuantumToClassicalPanel}
\end{figure*}
model
We now briefly outline the derivation of the classical model (all details are given in Appendix \ref{appendix:QuantumToClassical}). We use the Nakajima-Zwanzig approach \cite{Nakajima-PTP-1958,Zwanzig-JCP-1960} as in Ref.~\cite{Mitchison-New-J-Phys-2015}, introducing the following projection superoperators
\begin{equation}
    \mathcal{P}\hat{\rho} \equiv \sum_{j=0,L,R} \mel{j}{\hat{\rho}}{j}\dyad{j}, \hspace{1cm} \mathcal{Q}\equiv 1-\mathcal{P},
    \label{eq:NakajimaProjectors}
\end{equation}
which single out the diagonal ($\mathcal{P}$) and off-diagonal ($\mathcal{Q}$) elements of a density matrix $\hat{\rho}$ in the basis $\{\ket{0},\ket{L},\ket{R}\}$. With the superoperators, Eq.~(\ref{eq:QFPME_SepTS_D2}) can be separated into two coupled differential equations for $\mathcal{P}\hat{\rho}_t(D_1)$ and $\mathcal{Q}\hat{\rho}_t(D_1)$. The equation for $\mathcal{Q}\hat{\rho}_t(D_1)$ can be solved in terms of $\mathcal{P}\hat{\rho}_t(D_1)$, resulting in a closed equation of motion for $\mathcal{P}\hat{\rho}_t(D_1)$. Under strong dephasing $\Tilde{\Gamma}\gg\Gamma,\xi_2,\xi_3,\gamma_1$ ($\xi_{2/3}$ defined below), this equation can be approximated by the following classical Fokker-Planck master equation
\begin{widetext}
\begin{equation}
\begin{aligned}
\partial_t\boldsymbol{\rho}_t(D_1) = 
    &\begin{pmatrix}
    -\gamma_L(\epsilon_0)-\gamma_R(\epsilon_u) & \Tilde{\kappa}_L(D_1) & \Tilde{\kappa}_R(D_1) \\
    \gamma_L(\epsilon_0) & -\Tilde{\kappa}_L(D_1)-\Tilde{\xi}(D_1) & \Tilde{\xi}(D_1) \\
    \gamma_R(\epsilon_u) & \Tilde{\xi}(D_1) & \Tilde{\kappa}_R(D_1)-\Tilde{\xi}(D_1)
    \end{pmatrix}
    \boldsymbol{\rho}_t(D_1) \\ &\hspace{4cm}+ \gamma_1\partial_{D_1}
    \begin{pmatrix}
    D_1 & 0 & 0 \\
    0 & D_1+1 & 0 \\
    0 & 0 & D_1-1
    \end{pmatrix}
    \boldsymbol{\rho}_t(D_1) + \frac{\gamma_1^2}{8\lambda_1}\partial_{D_1}^2\boldsymbol{\rho}_t(D_1),
\end{aligned}
\label{eq:ClassicalQFPME}
\end{equation}
\end{widetext}
where we have vectorized $\mathcal{P}\hat{\rho}_t(D_1)\to[\mel{0}{\hat{\rho}_t(D_1)}{0},\mel{L}{\hat{\rho}_t(D_1)}{L},\mel{R}{\hat{\rho}_t(D_1)}{R}]^{\rm T}\equiv\boldsymbol{\rho}_t(D_1)$, and introduced the $D$-dependent rates
\begin{equation}
\begin{aligned}
        \Tilde{\kappa}_L(D_1) &= [1-\theta(D_1)]\kappa_L(\epsilon_d) + \theta(D_1)\kappa_L(\epsilon_u), \\
        \Tilde{\kappa}_R(D_1), &= [1-\theta(D_1)]\kappa_R(\epsilon_d) + \theta(D_1)\kappa_R(\epsilon_0), \\
        \Tilde{\xi}(D_1) &= [1-\theta(D_1)]\xi_2 + \theta(D_1)\xi_3,
\end{aligned}
\end{equation}
with the classical interdot tunneling rates
\begin{equation}
    \xi_2 = \frac{4g^2}{\kappa_L(\epsilon_d)+\kappa_R(\epsilon_d)+4\Tilde{\Gamma}}
\end{equation}
for configuration 2, and
\begin{equation}
    \xi_3 = \frac{4g^2[\kappa_L(\epsilon_u)+\kappa_R(\epsilon_0)+4\Tilde{\Gamma}]}{[\kappa_L(\epsilon_u)+\kappa_R(\epsilon_0)+4\Tilde{\Gamma}]^2+4(\epsilon_u-\epsilon_0)^2}
\end{equation}
for configuration 3. Due to the finite lifetime broadening of the quantum dots, these rates have a Lorentzian profile in the level detuning $\epsilon_L-\epsilon_R$ \cite{Sprekeler-PRB-2004,Kiesslich-PRB-2006,Kalaee-PRE-2021}. Therefore, for large detuning in configuation 3, interdot tunneling is suppressed. The rates are also suppressed for large dephasing $\Tilde{\Gamma}$, and vanish as $\Tilde{\Gamma}\to\infty$, reflecting that tunneling is a quantum phenomenon relying on the presence of coherence. However, the model shows that tunneling is well-modelled by classical jumps when the coherence is heavily suppressed, as illustrated in Fig.~\ref{fig:QtoCTrajectories}(b).

For a fast detector ($\gamma_1\gg\Gamma,\xi_2,\xi_3$) and a strong measurement ($\lambda_1\gg\gamma_1$), as well as large level detuning $|\epsilon_{u}-\epsilon_0|$ and $|\epsilon_{u/d}-\mu_{L/R}|\gg T$ in configurations 1 and 3 (such that $\xi_3\to0$), we find the following classical rate equation for the DQD (derivation Appendix \ref{appendix:QuantumToClassical}),
\begin{equation}
    \partial_t\boldsymbol{\rho}_t  = 
    \begin{pmatrix}
        -\gamma_L(\epsilon_0) & 0 & \kappa_R(\epsilon_0) \\
        \gamma_L(\epsilon_0) & -\xi_2 & 0 \\
        0 & \xi_2 & -\kappa_R(\epsilon_0)
    \end{pmatrix}
    \boldsymbol{\rho}_t,
\label{eq:ClassicalIdealModel}
\end{equation}
where $\boldsymbol{\rho}_t=\int dD_1 \boldsymbol{\rho}_t(D_1)$. Equation (\ref{eq:ClassicalIdealModel}) exactly coincides with the phenomenological model used in Ref.~\cite{Annby-PRB-2020} to study the classical operation of the demon. We thus provide a derivation for the phenomenological model from an underlying quantum description. Note that the rate for configuration 2 ($\xi_2$) is independent on the detuning $\epsilon_u-\epsilon_0$, implying that the phase damping effect highlighted in Fig.~\ref{fig:performance_simplifying_assumptions}(c) is not present in the classical model.% when studying fast detectors and taking the limit $|\epsilon_u-\epsilon_0|\gg T$. 
The reason for this is, that for the classical model, we assume $\tilde{\Gamma}\gg \gamma_1$, whereas the fast detector limit discussed in Sec.~\ref{sec:fast_detectors} requires $\gamma_1\gg\tilde{\Gamma}$. For details, see Appendix \ref{appendix:QuantumToClassical}.

To compare the quantum (\ref{eq:QFPME_SepTS_D2}) and classical (\ref{eq:ClassicalQFPME}) models, we plot the power, heat current and demon energy current as functions of $\lambda_1$ in Fig.~\ref{fig:QuantumToClassicalPanel}. The figure highlights that the system is well-described by the classical model for strong measurements, where quantum coherence is heavily suppressed. This is in accordance with the regime of validity of the classical model (see conditions above).

We note that Figs.~\ref{fig:QuantumToClassicalPanel}(a) and (b) illustrate that the classical model breaks down around $\lambda_1/\gamma_1\sim10^{-1}$, where the backaction of the measurement reduces. In this regime, we refer to the demon as being quantum, as the classical model fails to accurately describe the dynamics. Additionally, (a) and (b) identify parameter regimes where the respective models (quantum and classical) predict higher power than the other. This illustrates that quantum coherence can be neither seen as strictly detrimental or strictly beneficial in extracting work. We find the quantum model to predict higher power for smaller bandwidth of detector 1 ($\gamma_1$), while a larger bandwidth results in higher power in the classical model. This can be understood by the phase-damping effect discussed above, which is absent in the classical model and impedes power generation for a fast detector in the quantum model. We note that the demon is not generally energy conserving (per definition of Sec.~\ref{sec:energy_conserving}), except for large $\epsilon_u$ and large measurement strength, see panel (c). This is because for quantum effects to be relevant, measurement errors need to be allowed for. An energy conserving demon in the quantum regime can thus only be obtained by tailoring the tunnel rates as discussed in Sec.~\ref{sec:energy_conserving}. For strong measurements we find $\dot{Q}>0$ and $\dot{E}_{\rm D}<0$ for panels (a) and (b), illustrating that the demon is cooling the reservoirs by extracting work through the gates even when the Zeno effect blocks transport. Due to the choice of $\epsilon_u$ (and $\epsilon_d$), electrons can tunnel into the DQD in configuration 1 at $\epsilon_u$ from the right, or at $\epsilon_0$ from the left. If the left dot is occupied, the demon adjusts the levels to configuration 2, and if the right dot is occupied to configuration 3. As the strong measurement suppresses interdot tunneling, the electron can only tunnel out to the reservoir it entered from, but at a lower energy. A trajectory illustrating this is shown in Fig.~\ref{fig:performance_simplifying_assumptions}(e), where an electron tunnels into the left dot in configuration 1 and out to the left reservoir in configuration 2.

\begin{figure}[ht]
    \centering
    \includegraphics[scale=1]{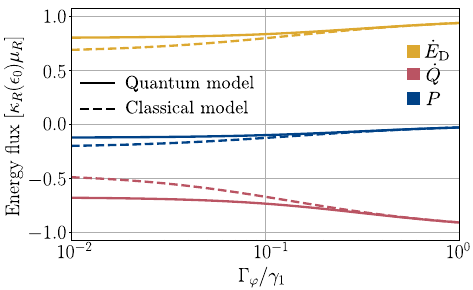}
    \caption{Effect of external dephasing. We plot the power (\ref{eq:EnsemblePower}), heat current (\ref{eq:EnsembleHeatCurrent}) and demon power (\ref{eq:EnsembleDemonEnergy}) as functions of $\Gamma_\varphi$. All lines were calculated numerically (App.~\ref{appendix:NumericalMethod}) using Eqs.~(\ref{eq:QFPME_SepTS_D2}) and (\ref{eq:ClassicalQFPME}) for the quantum and classical models, respectively. Solid lines correspond to the quantum model, dashed lines to the classical model. Under strong dephasing, the dynamics are well-described by the classical model. Parameters: $\Gamma/T=0.1$, $g/T=0.1$, $\epsilon_u/T=5$, $\epsilon_0/T=0$, $(\mu_R-\mu_L)/T=3$, $\lambda_1/T=1$, $\epsilon_d=-\epsilon_u$.}
    \label{fig:DephasingPlot}
\end{figure}

Finally, we illustrate the effect of the environmental dephasing rate $\Gamma_\varphi$ in Fig.~\ref{fig:DephasingPlot}. As $\Gamma_\varphi$ increases, the quantum and classical models merge as expected. %This is in accordance with the regime of validity of the classical model.

\section{Conclusion and outlook}
\label{sec:conclusions}

We studied an implementation of Maxwell's demon in a double quantum dot across the quantum-to-classical transition . We found a rich behavior as a function of the measurement strength. For strong measurements, or dephasing, the dynamics of the double dot is well-described by a classical model. In the limit of infinitely strong measurements, we observed the quantum Zeno effect, quenching the information-to-work conversion. For intermediate measurement strengths, the coherence of the DQD is preserved, allowing for information-to-work conversion under quantum dynamics. For weak measurements, we observed a Zeno-like effect, where measurement errors lead to random noise in the on-site energies of the dots, dephasing the system.
We anticipate our results to remain qualitatively valid for general quantum systems under continuous measurement and feedback control. Our results thus contribute to a better understanding of the interplay between information processing and coherence in continuously monitored and feedback-controlled devices and may help to guide implementations in quantum technology and quantum feedback control.

Future studies could optimize the feedback protocol for maximizing information-to-work conversion or quantum coherence in experimentally relevant scenarios. In the present work, we considered a Markovian description with an ideal total charge measurement. Relaxing these assumptions may result in interesting non-Markovian effects or uncover limitations in the presence of nonideal detectors. Furthermore, interesting nonclassical effects could be found by investigating fluctuations in the extracted work. This could be enabled by a full thermodynamic analysis relating the dynamics on the trajectory level with the QFPME.

\section*{Acknowledgments}
We acknowledge fruitful discussions with V. Maisi and K. Ensslin. This research was supported by FQXi Grant No.~FQXi-IAF19-07 from the Foundational Questions Institute Fund, a donor advised fund of Silicon Valley Community Foundation. P.B.~also acknowledges funding from the European Research Council (Consolidator grant ‘Cocoquest’ No. 101043705).
P.P.P. acknowledges funding from the Swiss National Science Foundation (Eccellenza Professorial Fellowship PCEFP2\_194268). P.S.~and B.A.A.~were supported by the Swedish Research Council, Grant No.~2018-03921. 
D.B., G.D.S.~and C.J.~acknowledge support from the John Templeton Foundation (award no.~62422).
D.B. acknowledges the University of Maryland supercomputing resources made available for conducting the research reported in this paper.

\newpage
\widetext
\appendix

\section*{Appendices}

\section{Justification of the local approach}
\label{appendix:LocalApproach}
In configuration 1 and 3, the protocol relies on the detuning $\epsilon_{u}-\epsilon_0$ being large enough to avoid unwanted tunneling between the dots, as well as between dots and reservoirs. However, for large detuning, the local master equation we employ in Eq.~\eqref{eq:DQDDynamics} may not be justified, see Ref.~\cite{Potts-NJP-2021}. Indeed, in this case the local model may be thermodynamically inconsistent, as energy appears to be created or destroyed during interdot transitions \cite{Levy-EPL-2014}. Note that these complications are absent in configuration 2, where the local approach results in a thermodynamically consistent description, as energy is conserved during interdot transitions, that is fully justified at small enough tunnel couplings \cite{Potts-NJP-2021}. In this appendix, we justify the local description for configurations 1 and 3 in Eq.~(\ref{eq:DQDDynamics}). We address this by noting that at sufficiently large detuning, a global master equation, ensuring thermodynamic consistency, is justified. Therefore, we consider here a global master equation for configurations 1 and 3, and compare to the results of the equations given in the main text, see Fig.~\ref{fig:performance_simplifying_assumptions}(b). For simplicity, we focus on the fast detector regime. We find that the local approach is justified as long as the coupling $g$ is sufficiently weak compared to the detuning. In this regime, inter-dot tunneling is blocked in configurations 1 and 3 in both the local and the global approaches.

The DQD Hamiltonian (\ref{eq:DQDHamiltonian}) can be diagonalized as $\hat{H}(\epsilon_L,\epsilon_R)=\sum_{j=0,1,2}E_j\dyad{E_j}$, with eigenenergies
\begin{equation}
    E_0 = 0, \hspace{2cm} E_{1/2} = \Bar{\epsilon} \pm \sqrt{\Delta^2+g^2},
\end{equation}
where $\Bar{\epsilon}=(\epsilon_L+\epsilon_R)/2$ and $\Delta=(\epsilon_L-\epsilon_R)/2$. The corresponding eigenvectors are
\begin{equation}
    \ket{E_0} = \ket{0}, \hspace{1cm} \ket{E_1} = a\ket{L}+b\ket{R}, \hspace{1cm} \ket{E_2} = c\ket{L} + d\ket{R},
\end{equation}
where
\begin{equation}
\begin{aligned}
&a = \frac{\Delta+\sqrt{g^2+\Delta^2}}{\sqrt{g^2+\left(\Delta+\sqrt{g^2+\Delta^2} \right)^2}}, \hspace{2cm} b=\frac{g}{\sqrt{g^2+\left(\Delta+\sqrt{g^2+\Delta^2} \right)^2}}, \\
&c = \frac{\Delta-\sqrt{g^2+\Delta^2}}{\sqrt{g^2+\left(\Delta-\sqrt{g^2+\Delta^2} \right)^2}}, \hspace{2cm} d = \frac{g}{\sqrt{g^2+\left(\Delta-\sqrt{g^2+\Delta^2} \right)^2}}.
\end{aligned}
\end{equation}
The gobal master equation is given by (see Ref.~\cite{Potts-NJP-2021} for a detailed derivation)
\begin{equation}
\begin{aligned}
    \partial_t\hat{\rho}_t = \mathcal{L}^{(g)}(\epsilon_L,\epsilon_R)\hat{\rho}_t = -i[\hat{H}(\epsilon_L,\epsilon_R),\hat{\rho}_t]  &+ \left[\gamma_L^{(a)}(E_1)+\gamma_R^{(b)}(E_1)\right]\mathcal{D}[\dyad{E_1}{E_0}]\hat{\rho}_t \\
    &+ \left[\gamma_L^{(c)}(E_2)+\gamma_R^{(d)}(E_2)\right]\mathcal{D}[\dyad{E_2}{E_0}]\hat{\rho}_t \\
    &+ \left[\kappa_L^{(a)}(E_1)+\kappa_R^{(b)}(E_1)\right]\mathcal{D}[\dyad{E_0}{E_1}]\hat{\rho}_t \\
    &+ \left[\kappa_L^{(c)}(E_2)+\kappa_R^{(d)}(E_2)\right]\mathcal{D}[\dyad{E_0}{E_2}]\hat{\rho}_t,
\end{aligned}
\end{equation}
where $\gamma_\alpha^{(j)}(E)=j^2\gamma_\alpha(E)$ and $\kappa_\alpha^{(j)}(E)=j^2\kappa_\alpha(E)$, with $j=a,b,c,d$. This equation is valid for $\sqrt{\Delta^2+g^2},\text{max}\{k_BT,|E_{1/2}-\mu_\alpha|\}\gg \Gamma_\alpha$ \cite{Potts-NJP-2021}. To include the global description in configuration 1 and 3, we use $\mathcal{L}_1^{(g)}=\mathcal{L}^{(g)}(\epsilon_0,\epsilon_u)$ and $\mathcal{L}_3^{(g)}=\mathcal{L}^{(g)}(\epsilon_u,\epsilon_0)$ in Eq.~(\ref{eq:FullFeedbackLiouvillian}). Under the exchange $(\epsilon_0,\epsilon_u)\to(\epsilon_u,\epsilon_0)$, we note that $\Delta\to-\Delta$, $E_j\to E_j$ for $j=0,1,2$, $a\to-c$, $b\to d$, $c\to-a$, and $d\to b$. Also note that $\ket{E_1}\to-c\ket{L}+d\ket{R}$ with eigenvalue $E_1$, and $\ket{E_2}\to-a\ket{L}+b\ket{R}$ with eigenvalue $E_2$ under this exchange. We are interested in the regime where $\Delta\gg g$, for which $E_1\simeq\epsilon_u$ and $E_2\simeq\epsilon_0$ to leading order in $g/\Delta\ll1$. Furthermore, $a$ and $c$ dominate over $b$ and $d$, such that the eigenstates are approximately the local states.

Here we calculate the power production by using full counting statistics \cite{schaller-book}. This means that we are interested in the probability distribution $P_t(\boldsymbol{n}) = \int d\boldsymbol{D}\trace\{\hat{\rho}_t(\boldsymbol{D},\boldsymbol{n})\}$ with $\boldsymbol{n}=(n_L^{(d)},n_R^{(d)},n_L^{(1)},n_R^{(1)},n_L^{(2)},n_R^{(2)})$ being a vector with the entries $n_\alpha^{(j)}$ denoting the number of electrons exchanged between dot $\alpha$ and bath $\alpha$ at energy $E_j$, where $E_d=\epsilon_d$, and $\hat{\rho}_t(\boldsymbol{D},\boldsymbol{n})$ is a $\boldsymbol{n}$-resolved density matrix. By introducing counting fields $\chi_\alpha^{(j)}$ for each $n_\alpha^{(j)}$, we define the density matrix $\hat{\rho}_t(\boldsymbol{D},\boldsymbol{\chi})=\sum_{\boldsymbol{n}}e^{i\boldsymbol{n}\cdot\boldsymbol{\chi}}\hat{\rho}_t(\boldsymbol{D},\boldsymbol{n})$, where $\boldsymbol{\chi}=(\chi_L^{(d)},\chi_R^{(d)},\chi_L^{(1)},\chi_R^{(1)},\chi_L^{(2)},\chi_R^{(2)})$. For this density matrix, Eq.~(\ref{eq:FullQFPME}) transforms into
\begin{equation}
    \partial_t\hat{\rho}_t(\boldsymbol{D},\boldsymbol{\chi}) = \mathcal{L}(\boldsymbol{D},\boldsymbol{\chi})\hat{\rho}_t(\boldsymbol{D},\boldsymbol{\chi}) + \Tilde{\Gamma}\mathcal{D}[\hat{A}_1]\hat{\rho}_t(\boldsymbol{D},\boldsymbol{\chi})
    + \sum_{j=1,2} \left[ \gamma_j \partial_{D_j} \mathcal{A}_j(D_j)\hat{\rho}_t(\boldsymbol{D},\boldsymbol{\chi}) + \frac{\gamma_j^2}{8\lambda_j} \partial_{D_j}^2 \hat{\rho}_t(\boldsymbol{D},\boldsymbol{\chi}) \right],
\end{equation}
with $ \mathcal{L}(\boldsymbol{D},\boldsymbol{\chi}) = \theta(D_2)\mathcal{L}_1^{(g)}(\boldsymbol{\chi}) + [1-\theta(D_1)][1-\theta(D_2)]\mathcal{L}_2(\boldsymbol{\chi}) + \theta(D_1)[1-\theta(D_2)]\mathcal{L}_3^{(g)}(\boldsymbol{\chi})$, where
\begin{equation}
    \begin{aligned}
        \mathcal{L}_3^{(g)}(\boldsymbol{\chi})\hat{\rho} = -i[\hat{H}(\epsilon_u,\epsilon_0),\hat{\rho}] &+ \gamma_L^{(a)}(E_1)\mathcal{D}^{(+)}_{\chi_L^{(1)}}[\dyad{E_1}{E_0}]\hat{\rho}+\gamma_R^{(b)}(E_1)\mathcal{D}^{(+)}_{\chi_R^{(1)}}[\dyad{E_1}{E_0}]\hat{\rho} \\
    &+ \gamma_L^{(c)}(E_2)\mathcal{D}^{(+)}_{\chi_L^{(2)}}[\dyad{E_2}{E_0}]\hat{\rho}+\gamma_R^{(d)}(E_2)\mathcal{D}^{(+)}_{\chi_R^{(2)}}[\dyad{E_2}{E_0}]\hat{\rho} \\
    &+ \kappa_L^{(a)}(E_1)\mathcal{D}^{(-)}_{\chi_L^{(1)}}[\dyad{E_0}{E_1}]\hat{\rho}+\kappa_R^{(b)}(E_1)\mathcal{D}^{(-)}_{\chi_R^{(1)}}[\dyad{E_0}{E_1}]\hat{\rho} \\
    &+ \kappa_L^{(c)}(E_2)\mathcal{D}^{(-)}_{\chi_L^{(2)}}[\dyad{E_0}{E_2}]\hat{\rho}+\kappa_R^{(d)}(E_2)\mathcal{D}^{(+)}_{\chi_R^{(2)}}[\dyad{E_0}{E_2}]\hat{\rho},
    \end{aligned}
\end{equation}
using that $\Delta=(\epsilon_u-\epsilon_0)/2$, and the counting field-resolved dissipator $\mathcal{D}_\chi^{(\pm)}[\hat{c}]\hat{\rho}=e^{\pm i\chi}\hat{c}\hat{\rho}\hat{c}^\dagger-\{\hat{c}^\dagger\hat{c},\hat{\rho}\}/2$, with $+(-)$ denoting whether an electron tunneled into (out of) a dot. Similarly, for configuration 2, we get
\begin{equation}
\begin{aligned}
    \mathcal{L}_2(\boldsymbol{\chi})\hat{\rho} = -i[\hat{H}(\epsilon_d,\epsilon_d),\hat{\rho}] &+ \gamma_L(\epsilon_d)\mathcal{D}_{\chi_L^{(d)}}^{(+)}[\dyad{L}{0}]\hat{\rho} + \kappa_L(\epsilon_d)\mathcal{D}_{\chi_L^{(d)}}^{(-)}[\dyad{0}{L}]\hat{\rho} \\
    &+\gamma_R(\epsilon_d)\mathcal{D}_{\chi_R^{(d)}}^{(+)}[\dyad{R}{0}]\hat{\rho} + \kappa_R(\epsilon_d)\mathcal{D}_{\chi_R^{(d)}}^{(-)}[\dyad{0}{R}]\hat{\rho}.
\end{aligned}
\end{equation}
To obtain the Liouvillian for configuration 1, we use $\mathcal{L}_3(\boldsymbol{\chi})$ with the transformation $\Delta\to-\Delta$, and get
\begin{equation}
    \begin{aligned}
        \mathcal{L}_1^{(g)}(\boldsymbol{\chi})\hat{\rho} = -i[\hat{H}(\epsilon_0,\epsilon_u),\hat{\rho}] &+ \gamma_L^{(c)}(E_1)\mathcal{D}^{(+)}_{\chi_L^{(1)}}[\dyad{E_1}{E_0}]\hat{\rho}+\gamma_R^{(d)}(E_1)\mathcal{D}^{(+)}_{\chi_R^{(1)}}[\dyad{E_1}{E_0}]\hat{\rho} \\
    &+ \gamma_L^{(a)}(E_2)\mathcal{D}^{(+)}_{\chi_L^{(2)}}[\dyad{E_2}{E_0}]\hat{\rho}+\gamma_R^{(b)}(E_2)\mathcal{D}^{(+)}_{\chi_R^{(2)}}[\dyad{E_2}{E_0}]\hat{\rho} \\
    &+ \kappa_L^{(c)}(E_1)\mathcal{D}^{(-)}_{\chi_L^{(1)}}[\dyad{E_0}{E_1}]\hat{\rho}+\kappa_R^{(d)}(E_1)\mathcal{D}^{(-)}_{\chi_R^{(1)}}[\dyad{E_0}{E_1}]\hat{\rho} \\
    &+ \kappa_L^{(a)}(E_2)\mathcal{D}^{(-)}_{\chi_L^{(2)}}[\dyad{E_0}{E_2}]\hat{\rho}+\kappa_R^{(b)}(E_2)\mathcal{D}^{(+)}_{\chi_R^{(2)}}[\dyad{E_0}{E_2}]\hat{\rho},
    \end{aligned}
\end{equation}
where the coefficients $a$, $b$, $c$, and $d$ are the same as for configuration 3, but $\ket{E_1}=-c\ket{L}+d\ket{R}$ and $\ket{E_2}=-a\ket{L}+b\ket{R}$. To carry out calculations in the $\{\ket{0},\ket{L},\ket{R}\}$-basis, the following relations are useful
\begin{equation}
\begin{aligned}
    \dyad{E_1}{E_0} &= a\hat{\sigma}_L^\dagger + b\hat{\sigma}_R^\dagger \to -c\hat{\sigma}_L^\dagger + d\hat{\sigma}_R^\dagger, \\
    \dyad{E_2}{E_0} &= c\hat{\sigma}_L^\dagger + d\hat{\sigma}_R^\dagger \to -a\hat{\sigma}_L^\dagger + b\hat{\sigma}_R^\dagger,
\end{aligned}
\end{equation}
where `$\to$' denotes the transformation under $\Delta\to-\Delta$, which should be used for configuration 1. Note that $\hat{\sigma}_\alpha=\dyad{0}{\alpha}$ for $\alpha=L,R$.

\begin{figure}
    \centering
    \includegraphics[scale=1]{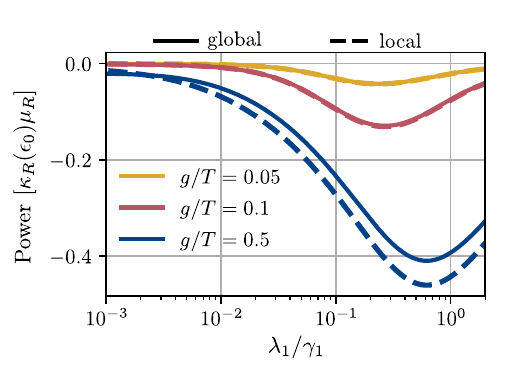}
    \caption{Comparing the local and global descriptions of configuration 1 and 3. We plot the power production as a function of $\lambda_1/\gamma_1$ for fast detectors (Sec.~\ref{sec:fast_detectors}). Solid lines correspond to Eq.~(\ref{eq:PowerLocalVsGlobal}), using the global description. The dashed lines are the same as in Fig.~\ref{fig:performance_simplifying_assumptions}(b), using the local description. For $\Delta\gg g$, the two descriptions overlap, justifying the local approach for configurations 1 and 3. As $g$ increases, the eigenenergies are no longer well-approximated by the local dot energies. This induces a discrepancy between the descriptions. Same parameters as in Fig.~\ref{fig:performance_simplifying_assumptions}(b).}
    \label{fig:PowerGlobalVSLocal}
\end{figure}

With the moment generating function
\begin{equation}
    \psi_t(\boldsymbol{\chi}) = \trace\{\hat{\rho}_t(\boldsymbol{\chi})\} = \int d\boldsymbol{D} \trace\{\hat{\rho}_t(\boldsymbol{D},\boldsymbol{\chi})\},
\end{equation}
we can find the average particle current between dot $\alpha$ and reservoir $\alpha$ at energy $j$ via \cite{Annby-PRL-2022}
\begin{equation}
    \langle \dot{n}_\alpha^{(j)} \rangle = -i\int d\boldsymbol{D}\trace\left\{\partial_{\chi_\alpha^{(j)}}\mathcal{L}(\boldsymbol{D},\boldsymbol{\chi})\hat{\rho}_t(\boldsymbol{D},\boldsymbol{\chi})\Big|_{\boldsymbol{\chi}=0}\right\}.
\end{equation}
For fast detectors (Sec.~\ref{sec:fast_detectors}), this reduces to
\begin{equation}
    \langle \dot{n}_\alpha^{(j)} \rangle = -i\trace\left\{\left[\partial_{\chi_\alpha^{(j)}}\mathcal{L}_{\rm fb}(\boldsymbol{\chi})\right]\Big|_{\boldsymbol{\chi}=0}\hat{\rho}_t \right\},
\end{equation}
where $\hat{\rho}_t=\int d\boldsymbol{D}\hat{\rho}_t(\boldsymbol{D})$ is the state of the DQD for fast detectors [Eq.~(\ref{eq:SepTSMasterEquation})], and
$\mathcal{L}_{\rm fb}(\boldsymbol{\chi}) = \mathcal{L}_1^{(g)}(\boldsymbol{\chi})\mathcal{V}_{00} + [(1-\eta)\mathcal{L}_2(\boldsymbol{\chi})+\eta\mathcal{L}_3^{(g)}(\boldsymbol{\chi})]\mathcal{V}_{LL}+[\eta\mathcal{L}_2(\boldsymbol{\chi})+(1-\eta)\mathcal{L}_3^{(g)}(\boldsymbol{\chi})]\mathcal{V}_{RR} + [\mathcal{L}_2(\boldsymbol{\chi})+\mathcal{L}_3^{(g)}(\boldsymbol{\chi})](\mathcal{V}_{LR}+\mathcal{V}_{RL})/2$. We can now compute the power as
\begin{equation}
    P = \sum_{\alpha=L,R}\sum_{j=d,1,2} \mu_\alpha \langle \dot{n}_\alpha^{(j)} \rangle.
    \label{eq:PowerLocalVsGlobal}
\end{equation}
This is plotted in Fig.~\ref{fig:PowerGlobalVSLocal} together with the power from Fig.~\ref{fig:performance_simplifying_assumptions}(b) for fast detectors, where configurations 1 and 3 were treated as local. We note that the global and local descriptions overlap for $g/T=0.05,0.1$, and lie close for $g/T=0.5$. In fact, as $g$ is increased, we expect the local and global descriptions to depart, as the approximations $E_1\simeq\epsilon_u$ and $E_2\simeq\epsilon_d$ break down. We conclude that the local descriptions of configuration 1 and 3 are valid as long as $\Delta\gg g$.

\section{Details on the energetics of the demon}
\label{appendix:DetailsEnergetics}

\begin{figure}
    \centering
    \includegraphics[scale=1]{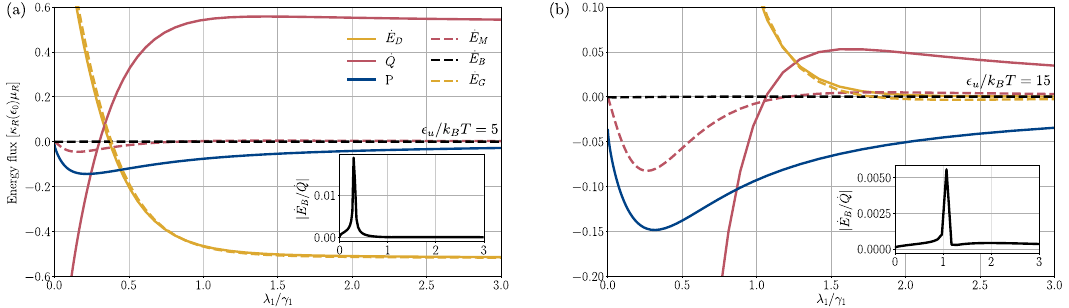}
    \caption{Energy fluxes as functions of $\lambda_1$. All lines were calculated numerically with Eq.~(\ref{eq:QFPME_SepTS_D2}). The inset shows the ratio $|\dot{E}_{\rm B}/\dot{Q}|$ as a function of $\lambda_1$. The figure illustrates that $\dot{E}_{\rm D}\simeq\dot{E}_{\rm G}$, and that $\dot{E}_{\rm B}$ is much smaller than the other fluxes. Parameters: $\gamma_1/k_BT=10$, $\Gamma/k_BT=0.1$, $g/k_BT=0.1$, $\Gamma_\varphi/k_BT=0$, $\epsilon_0/k_BT=0$, $(\mu_R-\mu_L)/k_BT=3$, $\epsilon_u=-\epsilon_d$, (a) $\epsilon_u/k_BT=5$, (b) $\epsilon_u/k_BT=15$.}
    \label{fig:AppendixEnergetics}
\end{figure}

In this appendix, we give the details of the calculations in Sec.~(\ref{sec:Energetics}). The average energy of the DQD is given by Eq.~(\ref{eq:AverageEnergyDQD}), where we can write the Hamiltonian as
\begin{equation}
    \hat{H}(\boldsymbol{D}) = \sum_{j=1}^3h_j(\boldsymbol{D})\hat{H}_j,
\end{equation}
where $\hat{H}_1=\hat{H}(\epsilon_0,\epsilon_u)$, $\hat{H}_2=\hat{H}(\epsilon_d,\epsilon_d)$, $\hat{H}_3=\hat{H}(\epsilon_u,\epsilon_0)$, and $h_j(\boldsymbol{D})$ is given in Eq.~(\ref{eq:FeedbackRegions}), satisfying $h_j(\boldsymbol{D})h_k(\boldsymbol{D})=\delta_{jk}h_j(\boldsymbol{D})$. The feedback-controlled Liouvillian can be written as
\begin{equation}
    \mathcal{L}(\boldsymbol{D})\hat{\rho} = \sum_{j=1}^3h_j(\boldsymbol{D})\left( -i[\hat{H}_j,\hat{\rho}] + \sum_{\alpha=L,R}\gamma_\alpha(\epsilon_\alpha^{(j)})\mathcal{D}[\hat{\sigma}_\alpha^\dagger]\hat{\rho} + \kappa_\alpha(\epsilon_\alpha^{(j)})\mathcal{D}[\hat{\sigma}_\alpha]\hat{\rho} \right),
\end{equation}
where we introduced the compact jump operator notation $\hat{\sigma}_\alpha=\dyad{0}{\alpha}$. Taking the time derivative of Eq.~(\ref{eq:AverageEnergyDQD}) gives
\begin{equation}
\begin{aligned}
    \dot{E}(t) = \int dD_1 dD_2 \trace\{\hat{H}(\boldsymbol{D})\mathcal{L}(\boldsymbol{D})\hat{\rho}_t(\boldsymbol{D})\} + \Tilde{\Gamma }\int dD_1 dD_2 \trace\{\hat{H}(\boldsymbol{D})\mathcal{D}[\hat{A}_1]\hat{\rho}_t(\boldsymbol{D})\} \\
    + \sum_{k=1}^2 \int dD_1dD_2 \trace\left\{\hat{H}(\boldsymbol{D}) \left[ \gamma_k\partial_{D_k}\mathcal{A}_k(D_k)+\frac{\gamma_k^2}{8\lambda_k}\partial_{D_k}^2\right]\hat{\rho}_t(\boldsymbol{D}) \right\},
\end{aligned}
\label{eq:AppendixTotalEnergyCurrent}
\end{equation}
where we used Eq.~(\ref{eq:FullQFPME}). The first term on the rhs can be manipulated as in the following series of equalities,
\begin{equation}
    \begin{aligned}
        &\int dD_1 dD_2 \trace\{\hat{H}(\boldsymbol{D})\mathcal{L}(\boldsymbol{D})\hat{\rho}_t(\boldsymbol{D})\} \\
        &= \int dD_1 dD_2 \sum_{j=1}^3 \sum_{\alpha=L,R} h_j(\boldsymbol{D}) \left( \gamma_\alpha(\epsilon_\alpha^{(j)})\trace\{\hat{H}_j\mathcal{D}[\hat{\sigma}_\alpha^\dagger]\hat{\rho}_t(\boldsymbol{D})\} + \kappa_\alpha(\epsilon_\alpha^{(j)})\trace\{\hat{H}_j\mathcal{D}[\hat{\sigma}_\alpha]\hat{\rho}_t(\boldsymbol{D})\} \right) \\
        &= \int dD_1 dD_2 \sum_{j=1}^3 \sum_{\alpha=L,R} h_j(\boldsymbol{D}) \Big( \gamma_\alpha(\epsilon_\alpha^{(j)})\trace\{(\hat{H}_j-\mu_\alpha\hat{N}_\alpha)\mathcal{D}[\hat{\sigma}_\alpha^\dagger]\hat{\rho}_t(\boldsymbol{D})\} + \kappa_\alpha(\epsilon_\alpha^{(j)})\trace\{(\hat{H}_j-\mu_\alpha\hat{N}_\alpha)\mathcal{D}[\hat{\sigma}_\alpha]\hat{\rho}_t(\boldsymbol{D})\} \\ &\hspace{5cm}+ \mu_\alpha\left[\gamma_\alpha(\epsilon_\alpha^{(j)})\trace\{\hat{N}_\alpha\mathcal{D}[\hat{\sigma}_\alpha^\dagger]\hat{\rho}_t(\boldsymbol{D})\} + \kappa_\alpha(\epsilon_\alpha^{(j)})\trace\{\hat{N}_\alpha\mathcal{D}[\hat{\sigma}_\alpha]\hat{\rho}_t(\boldsymbol{D})\} \right] \Big) \\
        &= \int dD_1dD_2 \sum_{j=1}^3 \sum_{\alpha=L,R} h_j(\boldsymbol{D}) \Big\{ (\epsilon_\alpha^{(j)}-\mu_\alpha)\left[\gamma_\alpha(\epsilon_\alpha^{(j)})\rho_{00}(\boldsymbol{D})-\kappa_\alpha(\epsilon_\alpha^{(j)})\rho_{\alpha\alpha}(\boldsymbol{D})\right] \\ &\hspace{5cm}+ \mu_\alpha\left[\gamma_\alpha(\epsilon_\alpha^{(j)})\rho_{00}(\boldsymbol{D})-\kappa_\alpha(\epsilon_\alpha^{(j)})\rho_{\alpha\alpha}(\boldsymbol{D})\right] -g\kappa_\alpha(\epsilon_\alpha^{(j)})\Re{\rho_{LR}(\boldsymbol{D})} \Big\},
    \end{aligned}
\end{equation}
where $\hat{N}_\alpha=\dyad{\alpha}$ is the number operator of dot $\alpha$, $\mu_\alpha$ is the chemical potential of bath $\alpha$, and $\rho_{ab}(\boldsymbol{D})=\mel{a}{\hat{\rho}_t(\boldsymbol{D})}{b}$. We can now do the following decomposition,
\begin{equation}
    \int dD_1 dD_2 \trace\{\hat{H}(\boldsymbol{D})\mathcal{L}(\boldsymbol{D})\hat{\rho}_t(\boldsymbol{D})\} = P + \dot{Q} + \dot{E}_{\rm B},
\end{equation}
where we have identified the power and heat current
\begin{equation}
    P=\sum_{\alpha=L,R}\sum_{j=1}^3 \mu_\alpha \langle \dot{n} _{\alpha}^{(j)}\rangle \hspace{2cm} \dot{Q}=\sum_{\alpha=L,R}\sum_{j=1}^3 (\epsilon_\alpha^{(j)}-\mu_\alpha) \langle \dot{n} _{\alpha}^{(j)}\rangle,
\end{equation}
and the particle current towards the DQD from bath $\alpha$ in level configuration $j$
\begin{equation}
    \langle\dot{n}_{\alpha}^{(j)}\rangle = \int dD_1dD_2 h_j(\boldsymbol{D})[\gamma_\alpha(\epsilon_\alpha^{(j)})\rho_{00}(\boldsymbol{D}) - \kappa_\alpha(\epsilon_\alpha^{(j)})\rho_{\alpha\alpha}(\boldsymbol{D}) ]
\end{equation}
as given in Eqs.~(\ref{eq:EnsemblePower})-(\ref{eq:ParticleCurrents}). The energy flux $\dot{E}_{\rm B}$ is defined as
\begin{equation}
    \dot{E}_{\rm B} = - \sum_{\alpha=L,R}\sum_{j=1}^3 g\kappa_\alpha(\epsilon_\alpha^{(j)}) \\ \int dD_1dD_2 h_j(\boldsymbol{D})\Re{\rho_{LR}(\boldsymbol{D})}.
\end{equation}
Within the local master equation approach this contribution is small and thus negligible in the thermodynamic book-keeping \cite{Potts-NJP-2021}. Further evidence supporting this claim is presented in Fig.~\ref{fig:AppendixEnergetics}, where we highlight that $\dot{E}_{\rm B}$ is significantly smaller than the remaining energy currents. Therefore, we put $\dot{E}_{\rm B}=0$.

The second term of Eq.~(\ref{eq:AppendixTotalEnergyCurrent}) is due to measurement backaction and environmental noise, and is given by
\begin{equation}
        \dot{E}_{\rm M} = \Tilde{\Gamma}\int dD_1dD_2 \trace\{\hat{H}(\boldsymbol{D})\mathcal{D}[\hat{A}_1]\hat{\rho}_t(\boldsymbol{D})\} = - 4g\Tilde{\Gamma}\Re{\rho_{LR}},
\end{equation}
where $\rho_{LR}=\int dD_1dD_2\mel{L}{\hat{\rho}_t(\boldsymbol{D}))}{R}$.

The final term of Eq.~(\ref{eq:AppendixTotalEnergyCurrent}) corresponds to the power performed by the voltage gates, and can be identified as
\begin{equation}
    \dot{E}_{\rm G} = \sum_{k=1,2} \int dD_1dD_2 \trace\Big\{\hat{H}(\boldsymbol{D}) \Big[ \gamma_k \partial_{D_k} \mathcal{A}_k(D_k)  + \frac{\gamma_k^2}{8\lambda_k} \partial_{D_k}^2 \Big]\hat{\rho}_t(\boldsymbol{D})\Big\}.
\end{equation}
Together with $E_{\rm M}$, we define the demon energy $\dot{E}_{\rm D}=\dot{E}_{\rm M}+\dot{E}_{\rm G}$, corresponding to all energy contributions provided by the demon.

The time derivative of Eq.~(\ref{eq:AverageEnergyDQD}) can now be written as
\begin{equation}
    \dot{E}(t) = P + \dot{Q} + \dot{E}_{\rm D}.
\end{equation}
Due to the discontinuities in $\hat{H}(\boldsymbol{D})$ and $\mathcal{L}(\boldsymbol{D})$, it is not straightforward to evaluate the integrals in $\dot{E}_{\rm D}$ (recall the definition of $\dot{E}_{\rm G}$). Computation techniques for these integrals are left for future work. However, in the stationary state, we infer the energy flux of the demon via
\begin{equation}
    \dot{E}_{\rm D} = -(P+\dot{Q}).
\end{equation}
Similarly, the gate power can be found via $\dot{E}_{\rm G}=-(P+\dot{Q}+\dot{E}_{\rm M})$. 

In Fig.~\ref{fig:AppendixEnergetics}, we plot $P$, $\dot{Q}$, $\dot{E}_{\rm D}$, $\dot{E}_{\rm B}$, $\dot{E}_{\rm M}$, and $\dot{E}_{\rm G}$ as functions of $\lambda_1$. All lines were calculated numerically with Eq.~(\ref{eq:QFPME_SepTS_D2}). The figure (see the inset) provides evidence that $\dot{E}_{\rm B}$ is much smaller than the other energy fluxes, and it is justified to neglect this quantity from the energetic book-keeping. Additionally, the plots highlight that $\dot{E}_{\rm D}\simeq\dot{E}_{\rm G}$, showing that the gate work is the main energetic contribution of the demon.

\section{Numerical method}
\label{appendix:NumericalMethod}

In this appendix, we detail the numerical method for solving Eq.~(\ref{eq:QFPME_SepTS_D2}). We begin by expanding the density matrix elements $\rho_{aa'}(D_1)=\mel*{a}{\hat{\rho}_t(D_1)}{a'}$, with $\ket{a}$ and $\ket*{a'}$ being states in the local basis $\{\ket{0},\ket{L},\ket{R}\}$), in terms of generalized Hermite polynomials as
\begin{equation}
    \rho_{aa'}(D_1) = \sum_{n=0}^{N-1}c_n^{(aa')}\frac{He_n^{[\sigma]}(D_1)}{\sqrt{n!\sigma^n}}\frac{e^{-D_1^2/2\sigma}}{\sqrt{2\pi\sigma}},
    \label{eq:AppendixExpansionOfRho}
\end{equation}
where the sum has been truncated at $N$ terms (for all simulations, we used $N=100$), $c_n^{(aa')}$ are expansion coefficients, $\sigma=\gamma_1/8\lambda_1$, and
\begin{equation}
    He_n^{[\sigma]}(D_1) = \left( \frac{\sigma}{2}\right)^{n/2} H_n\left( \frac{D_1}{\sqrt{2\sigma}} \right)
\end{equation}
are generalized Hermite polynomials of variance $\sigma$, with $H_n(D_1)=(-1)^ne^{D_1^2}\partial_{D_1}^ne^{-D_1^2}$ being the standard physicist's Hermite polynomials. By plugging this expansion of $\hat{\rho}_t(D_1)$ into Eq.~(\ref{eq:QFPME_SepTS_D2}), multiplying the equation with $He_m^{[\sigma]}(D_1)/\sqrt{m!\sigma^m}$, and integrating over $D_1$, we obtain a system of coupled recursion relations for the expansion coefficients $c_m^{(aa')}$. To find these relations, it is useful to note that
\begin{equation}
    \int_{-\infty}^\infty dD_1 \frac{He_m^{[\sigma]}(D_1)}{\sqrt{m!\sigma^m}} \frac{He_n^{[\sigma]}(D_1)}{\sqrt{n!\sigma^n}}\frac{e^{-D_1^2/2\sigma}}{\sqrt{2\pi\sigma}} = \delta_{nm},
\end{equation}
and
\begin{equation}
I_{mn} = \frac{1}{\sqrt{2 \pi \sigma^{n+m+1} n! m!}} \int_{0}^{\infty} He_m^{[\sigma]}(D_1) He_n^{[\sigma]}(D_1) e^{-D_1^2/2\sigma} dD_1 = 
\begin{cases}
1/2, \hspace{0.5cm} n=m \\
0, \hspace{0.5cm} n+m \hspace{0.1cm} \text{even}, n\neq m\\
C_{nm}, \hspace{0.5cm} n+m \hspace{0.1cm} \text{odd},
\end{cases}
\label{eq:AppendixIntegralPosD}
\end{equation}
where
\begin{equation}
C_{nm} = \frac{(-1)^{(n+m-1)/2}m!!(n-1)!!}{\sqrt{2 \pi n! m!}(m-n)}
\label{eq:AppendixCoefficient}
\end{equation}
for even $n$ and odd $m$. Note that the double factorial $(-1)!!=1$. Furthermore, after integrating over $D_1$, it is useful to note that the following replacements of terms can be done in Eq.~(\ref{eq:QFPME_SepTS_D2}),
\begin{itemize}
\item $\rho_{aa'}(D_1) \rightarrow c_m^{(aa')}$,
\item $\theta(D)\rho_{aa'}(D_1) \rightarrow \sum_n I_{mn} c_n^{(aa')}$
\item $\gamma \left[ \partial_{D_1}(D_1+k)\rho_{aa'}(D_1) + \sigma \partial_{D_1}^2 \rho_{aa'}(D_1) \right] \rightarrow -\gamma \left[ mc_m^{(aa')} + k \sqrt{\frac{m}{\sigma}} c_{m-1}^{(aa')} \right]$.
\end{itemize}
By forming vectors $\boldsymbol{c}_{aa'}=(c_0^{(aa')},c_1^{(aa')},\dots,c_{N-1}^{(aa')})^{\rm T}$, Eq.~(\ref{eq:QFPME_SepTS_D2}) can be written as the following matrix equation,
\begin{equation}
    \partial_t 
    \begin{pmatrix}
        \boldsymbol{c}_{00} \\ \boldsymbol{c}_{LL} \\ \boldsymbol{c}_{RR} \\ \boldsymbol{c}_{LR} \\ \boldsymbol{c}_{RL}
    \end{pmatrix}
    = M 
    \begin{pmatrix}
        \boldsymbol{c}_{00} \\ \boldsymbol{c}_{LL} \\ \boldsymbol{c}_{RR} \\ \boldsymbol{c}_{LR} \\ \boldsymbol{c}_{RL}
    \end{pmatrix},
\end{equation}
where
\begin{equation}
M = 
\begin{pmatrix}
-[\gamma_L(\epsilon_0)+\gamma_R(\epsilon_u)]\mathbb{1} + F^{(a)} & X_L+\xi_LJ & X_R+\xi_RJ & 0 & 0 \\
\gamma_L(\epsilon_0)\mathbb{1} &  -X_L-\xi_LJ + F^{(b)} & 0 & ig\mathbb{1} & -ig\mathbb{1} \\
\gamma_R(\epsilon_u)\mathbb{1} & 0 & -X_R-\xi_RJ+F^{(c)} & -ig\mathbb{1} & ig\mathbb{1} \\
0 & ig\mathbb{1} & -ig\mathbb{1} & Y+\mu_2J+F^{(a)} & 0 \\
0 & -ig\mathbb{1} & ig\mathbb{1} & 0 & Y^*+\mu_2^*J+F^{(a)}
\end{pmatrix}.
\end{equation}
with $\mathbb{1}$ being the identity matrix in $N$ dimensions, $X_L = [\kappa_L(\epsilon_u)+\kappa_L(\epsilon_d)]\mathbb{1}/2$, $X_R = [\kappa_R(\epsilon_0)+\kappa_R(\epsilon_d)]\mathbb{1}/2$, $Y=(\mu_2/2 - \mu_1)\mathbb{1}$,
\begin{equation}
J =
\begin{pmatrix}
0 & I_{01} & 0 & I_{03} & \cdots \\
I_{10} & 0 & I_{12} & 0 & \cdots \\
0 & I_{21} & 0 & I_{23} & \cdots \\
\vdots & \vdots & \vdots & \ddots 
\end{pmatrix},
\end{equation}
with $I_{mn}$ being defined in Eq.~(\ref{eq:AppendixIntegralPosD}),
\begin{equation}
F^{(a)} = - \gamma_1 
\begin{pmatrix}
0 & 0 & 0 & \cdots & 0 \\
0 & 1 & 0 & \cdots & 0 \\
0 & 0 & 2 & \hdots & 0 \\
\vdots & \vdots & \vdots & \ddots \\
0 & 0 & 0 & \cdots & N-1 
\end{pmatrix},
\end{equation}
\begin{equation}
F^{(b)} = - \gamma_1 
\begin{pmatrix}
0 & 0 & 0 & \cdots & 0 & 0 \\
\sqrt{1/\sigma_1} & 1 & 0 & \cdots & 0 & 0 \\
0 & \sqrt{2/\sigma_1} & 2 & \hdots & 0 & 0 \\
\vdots & \vdots & \vdots &  & \vdots & \vdots \\
0 & 0 & 0 & \cdots & \sqrt{(N-1)/\sigma_1} & N-1 
\end{pmatrix}
\end{equation}
and
\begin{equation}
F^{(c)} = - \gamma_1 
\begin{pmatrix}
0 & 0 & 0 & \cdots & 0 & 0 \\
-\sqrt{1/\sigma_1} & 1 & 0 & \cdots & 0 & 0 \\
0 & -\sqrt{2/\sigma_1} & 2 & \hdots & 0 & 0 \\
\vdots & \vdots & \vdots &  & \vdots & \vdots \\
0 & 0 & 0 & \cdots & -\sqrt{(N-1)/\sigma_1} & N-1 
\end{pmatrix}.
\end{equation}
In the above expressions, we introduced the parameters
\begin{equation}
\begin{aligned}
\xi_L &= \kappa_L(\epsilon_u)- \kappa_L(\epsilon_d), \\
\xi_R &= \kappa_R(\epsilon_0)- \kappa_R(\epsilon_d), \\
\mu_1 &= 2\tilde{\Gamma} + [\kappa_L(\epsilon_d)+\kappa_R(\epsilon_d)]/2, \\
\mu_2 &= [2i(\epsilon_0-\epsilon_u)+\kappa_L(\epsilon_d)+\kappa_R(\epsilon_d)-\kappa_L(\epsilon_u)-\kappa_R(\epsilon_0)]/2.
\end{aligned}
\end{equation}
The stationary state of Eq.~(\ref{eq:QFPME_SepTS_D2}) is found by numerically identifying the null space of $M$, giving the expansion coefficients of Eq.~(\ref{eq:AppendixExpansionOfRho}).

To calculate the particle currents in Eq.~(\ref{eq:ParticleCurrents}), it is useful to apply the following identities,
\begin{equation}
\begin{aligned}
\int_0^{\infty} dD_1 \rho_{aa'}(D_1) &= \frac{c^{(aa')}_0}{2} + \sum_{m \hspace{1mm} \text{odd}} C_{0m} c^{(aa')}_m, \\
\int_{-\infty}^{0} dD_1 \rho_{aa'}(D_1) &= \frac{c^{(aa')}_0}{2} - \sum_{m \hspace{1mm} \text{odd}} C_{0m} c^{(aa')}_m,
\end{aligned}
\end{equation}
where $C_{nm}$ is defined in Eq.~(\ref{eq:AppendixCoefficient}).

\section{Fast detectors}
\label{appendix:FastDetectors}

In this appendix, we derive Eqs.~(\ref{eq:QFPME_SepTS_D2}) and (\ref{eq:SepTSMasterEquation}), starting from Eq.~(\ref{eq:FullQFPME}). Both derivations rely on the assumption that the detectors are fast. To begin, we write the Quantum Fokker-Planck master equation (\ref{eq:FullQFPME}) as
\begin{equation}
    \partial_t \hat{\rho}_t(\boldsymbol{D}) = \mathcal{L}(\boldsymbol{D})\hat{\rho}_t(\boldsymbol{D}) + \Tilde{\Gamma}\mathcal{D}[\hat{A}_1]\hat{\rho}_t(\boldsymbol{D}) + \sum_{j=1,2} \mathcal{F}_j(D_j)\hat{\rho}_t(\boldsymbol{D}),
    \label{eq:AppendixFullQFPME}
\end{equation}
where we introduced the compact superoperator notation
\begin{equation}
    \mathcal{F}_j(D_j) = \gamma_j\partial_{D_j}\mathcal{A}_j(D_j) + \frac{\gamma_j^2}{8\lambda_j}\partial_{D_j}^2.
    \label{eq:AppendixFokkePlanckSupOp}
\end{equation}
To derive Eq.~(\ref{eq:QFPME_SepTS_D2}), we restrict ourselves to the regime where detector 2 is fast, i.e., when $\gamma_2\gg\text{max}\{\Gamma,g,|\epsilon_{u/d}-\epsilon_0|,\Tilde{\Gamma},\gamma_1\}$. Under this hierarchy of timescales, the density matrix $\hat{\rho}_t(\boldsymbol{D})$ can be expanded in orders of $1/\gamma_2$. Following the results of Ref.~\cite{Annby-PRL-2022} (see Sec.~II of the supplemental material of Ref.~\cite{Annby-PRL-2022}), the density matrix can, to zeroth order in $1/\gamma_2$, be written as
\begin{equation}
    \hat{\rho}_t(\boldsymbol{D}) = \sum_{aa'} \pi_{aa'}^{(2)}(D_2)\mathcal{V}_{aa'}\hat{\rho}_t(D_1),
    \label{eq:AppendixZerothOrderDensityMatrixD2}
\end{equation}
where the indices $a,a'\in\{0,L,R\}$, $\mathcal{V}_{aa'}\hat{\rho}=\mel{a}{\hat{\rho}}{a'}\dyad*{a}{a'}$, $\hat{\rho}_t(D_1)=\int dD_2\hat{\rho}_t(\boldsymbol{D})$, and
\begin{equation}
    \pi_{aa'}^{(j)}(D_j) = \sqrt{\frac{4\lambda_j}{\pi\gamma_j}}e^{-\frac{4\lambda_j}{\gamma_j}\left(D_j-\frac{\xi_a^{(j)}+\xi_{a'}^{(j)}}{2} \right)^2},
    \label{eq:AppendixDetector}
\end{equation}
where $j=1,2$ refers to the number of the detector, and $\hat{A}_j\ket{a}=\xi_{a}^{(j)}\ket{a}$. With Eq.~(\ref{eq:AppendixZerothOrderDensityMatrixD2}), $\mathcal{F}_2(D_2)\hat{\rho}_t(\boldsymbol{D})=0$. By plugging Eq.~(\ref{eq:AppendixZerothOrderDensityMatrixD2}) into Eq.~(\ref{eq:AppendixFullQFPME}) and integrating over $D_2$, the first term of Eq.~(\ref{eq:AppendixFullQFPME}) can be written as
\begin{equation}
\begin{aligned}
    &\int dD_2 \mathcal{L}(\boldsymbol{D})\hat{\rho}_t(\boldsymbol{D}) \\
    &\hspace{1cm}= \Bigg\{ \mathcal{L}_1 \Big[ (1-\eta_2)\mathcal{V}_{00} + \eta_2(1-\mathcal{V}_{00})\Big] + \Big[[1-\theta(D_1)]\mathcal{L}_2+\theta(D_1)\mathcal{L}_3\Big]\Big[ \eta_2\mathcal{V}_{00} + (1-\eta_2)(1-\mathcal{V}_{00}) \Big] \Bigg\} \hat{\rho}_t(D_1),
\end{aligned}
\end{equation}
where we used that $\mathcal{V}_{00}+\mathcal{V}_{LL}+\mathcal{V}_{RR}+\mathcal{V}_{LR}+\mathcal{V}_{RL} = 1$, and introduced the feedback error probability
\begin{equation}
    \eta_j = \frac{1}{2}\left[ 1-\erf\left(2\sqrt{\lambda_j/\gamma_j}\right) \right],
    \label{eq:AppendixErrorProb}
\end{equation}
 for detector $j$, with $\erf(\cdot)$ being the error function. In the limit $\lambda_2\to\infty$, i.e., under ideal total charge detection, we find that
\begin{equation}
    \int dD_2 \mathcal{L}(\boldsymbol{D})\hat{\rho}_t(\boldsymbol{D}) \\
    \to \Bigg\{ \mathcal{L}_1\mathcal{V}_{00} + \Big[[1-\theta(D_1)]\mathcal{L}_2+\theta(D_1)\mathcal{L}_3\Big](1-\mathcal{V}_{00}) \Bigg\} \hat{\rho}_t(D_1)\equiv\tilde{\mathcal{L}}(D_1)\hat{\rho}_t(D_1).
\end{equation}
For the second and third term of Eq.~(\ref{eq:AppendixFullQFPME}), the integration results in the replacement $\hat{\rho}_t(\boldsymbol{D})\to\hat{\rho}_t(D_1)$, and Eq.~(\ref{eq:FullQFPME}) reduces to Eq.~(\ref{eq:QFPME_SepTS_D2}). By vectorizing the density matrix as $\hat{\rho}_t(D_1)\to[\rho_{00}(D_1),\rho_{LL}(D_1),\rho_{RR}(D_1),\rho_{LR}(D_1),\rho_{RL}(D_1)]^{\rm T}$, with $\rho_{aa'}(D_1)=\mel{a}{\hat{\rho}_t(D_1)}{a'}$, we find the following matrix representations
\begin{equation}
\mathcal{L}_1\mathcal{V}_{00} =
\begin{pmatrix}
-\gamma_L(\epsilon_0)-\gamma_R(\epsilon_u) & 0 & 0 & 0 & 0 \\
\gamma_L(\epsilon_0) & 0 & 0 & 0 & 0 \\
\gamma_R(\epsilon_u) & 0 & 0 & 0 & 0 \\
0 & 0 & 0 & 0 & 0 \\
0 & 0 & 0 & 0 & 0
\end{pmatrix}
\end{equation}
for level configuration 1,
\begin{equation}
\mathcal{L}_2(1-\mathcal{V}_{00})=
\begin{pmatrix}
0 & \kappa_L(\epsilon_d) & \kappa_R(\epsilon_d) & 0 & 0 \\
0 & -\kappa_L(\epsilon_d) & 0 & ig & -ig \\
0 & 0 & -\kappa_R(\epsilon_d) & -ig & ig \\
0 & ig & -ig & \alpha_2 & 0 \\
0 & -ig & ig & 0 & \alpha_2
\end{pmatrix}
\label{eq:L_L_matrix_ideal_A2}
\end{equation}
for level configuration 2 with $\alpha_2=-[\kappa_L(\epsilon_d)+\kappa_R(\epsilon_d)]/2$, and
\begin{equation}
\mathcal{L}_3(1-\mathcal{V}_{00})=
\begin{pmatrix}
0 & \kappa_L(\epsilon_u) & \kappa_R(\epsilon_0) & 0 & 0 \\
0 & -\kappa_L(\epsilon_u) & 0 & ig & -ig \\
0 & 0 & -\kappa_R(\epsilon_0) & -ig & ig \\
0 & ig & -ig & \alpha_3 & 0 \\
0 & -ig & ig & 0 & \alpha_3^*
\end{pmatrix}
\end{equation}
for level configuration 3 with $\alpha_3 = -i(\epsilon_u-\epsilon_0) -[\kappa_L(\epsilon_u)+\kappa_R(\epsilon_0)]/2$. That is, due to the ideal total charge measurement, electrons can only enter the DQD in configuration 1, and only leave the DQD in configurations 2 and 3. 

We now derive Eq.~(\ref{eq:SepTSMasterEquation}), starting from Eq.~(\ref{eq:QFPME_SepTS_D2}). We begin by assuming that detector 1 is fast as well, i.e., $\gamma_2\gg\gamma_1\gg\text{max}\{\Gamma,g,|\epsilon_{u/d}-\epsilon_0|,\Tilde{\Gamma}\}$. In fact, it is equivalent to put $\gamma_1,\gamma_2\gg\text{max}\{\Gamma,g,|\epsilon_{u/d}-\epsilon_0|,\Tilde{\Gamma}\}$. Similar to the case above, we can write the density matrix in Eq.~(\ref{eq:QFPME_SepTS_D2}), to zeroth order in $1/\gamma_1$, as
\begin{equation}
    \hat{\rho}_t(D_1) = \sum_{aa'} \pi_{aa'}^{(1)}(D_1)\mathcal{V}_{aa'}\hat{\rho}_t,
    \label{eq:appendix:FastDetectorOne}
\end{equation}
where $\hat{\rho}_t=\int dD_1 \hat{\rho}_t(D_1)$, and $\pi_{aa'}^{(1)}(D_1)$ is given by Eq.~(\ref{eq:AppendixDetector}). By plugging this into Eq.~(\ref{eq:QFPME_SepTS_D2}), and integrating over $D_1$ yields Eq.~(\ref{eq:SepTSMasterEquation}), with the Liouvillian
\begin{equation}
    \mathcal{L}_{\rm fb} = \mathcal{L}_1\mathcal{V}_{00} + [(1-\eta_1)\mathcal{L}_2+\eta_1\mathcal{L}_3]\mathcal{V}_{LL}+[\eta_1\mathcal{L}_2+(1-\eta_1)\mathcal{L}_3]\mathcal{V}_{RR} + (\mathcal{L}_2+\mathcal{L}_3)(\mathcal{V}_{LR}+\mathcal{V}_{RL})/2,
\end{equation}
where the error probability $\eta_1$ is defined in Eq.~(\ref{eq:AppendixErrorProb}). In the main text, we put $\eta=\eta_1$. By vectorizing the density matrix $\hat{\rho}_t\to(\rho_{00},\rho_{LL},\rho_{RR},\rho_{LR},\rho_{RL})^{\rm T}$, the Liouvillian, and the term $\Tilde{\Gamma}\mathcal{D}[\hat{A}_1]$ [see Eq.~(\ref{eq:SepTSMasterEquation})], can be written in matrix form as
\begin{equation}
\mathcal{L}_{\rm fb} \to  
\begin{pmatrix}
-\gamma_L - \gamma_R & \kappa_L & \kappa_R & 0 & 0 \\
\gamma_L & -\kappa_L & 0 & ig & -ig \\
\gamma_R & 0 & -\kappa_R & -ig & ig \\
0 & ig & -ig & \alpha & 0 \\
0 & -ig & ig & 0 & \alpha^*
\end{pmatrix},
\hspace{2cm}
\Tilde{\Gamma}\mathcal{D}[\hat{A}_1] \to 
\begin{pmatrix}
    0 & 0 & 0 & 0 & 0 \\
    0 & 0 & 0 & 0 & 0 \\
    0 & 0 & 0 & 0 & 0 \\
    0 & 0 & 0 & -2\Tilde{\Gamma} & 0 \\
    0 & 0 & 0 & 0 & -2\Tilde{\Gamma} \\
\end{pmatrix},
\end{equation}
where
\begin{equation}
\begin{cases}
\gamma_L = \gamma_L(\epsilon_0), \\
\gamma_R = \gamma_R(\epsilon_u), \\
\kappa_L = (1-\eta)\kappa_L(\epsilon_d) + \eta\kappa_L(\epsilon_u), \\
\kappa_R = (1-\eta)\kappa_R(\epsilon_0) + \eta\kappa_R(\epsilon_d), \\
\alpha = -i \frac{\epsilon_u-\epsilon_0}{2} - \frac{\kappa_L(\epsilon_u) + \kappa_R(\epsilon_0) + \kappa_L(\epsilon_d) + \kappa_R(\epsilon_d)}{4}.
\end{cases}
\label{eq:AppendixSepTSRates}
\end{equation}
That is, after integrating out both detectors, the effect of feedback enters $\mathcal{L}_{\rm fb}$ by modifying the transition rates (\ref{eq:AppendixSepTSRates}) and introducing a factor of $1/2$ to the detuning $\epsilon_u-\epsilon_0$. Note that the individual level configurations no longer can be resolved. 

By using the fast detector assumptions [Eqs.~(\ref{eq:AppendixZerothOrderDensityMatrixD2}) and (\ref{eq:appendix:FastDetectorOne})] as well as the stationary solution of Eq.~(\ref{eq:SepTSMasterEquation}) in Eqs.~(\ref{eq:ParticleCurrents}), (\ref{eq:EnsemblePower}), and (\ref{eq:EnsembleHeatCurrent}), we find the expressions for power and heat in Eqs.~(\ref{eq:sepTSPower}) and (\ref{eq:sepTSHeatCurrent}).

\section{Quantum-to-classical transition}
\label{appendix:QuantumToClassical}

In this appendix, we derive Eqs~(\ref{eq:ClassicalQFPME}) and (\ref{eq:ClassicalIdealModel}). Our starting point is Eq.~(\ref{eq:QFPME_SepTS_D2}), here written as
\begin{equation}
    \partial_t\hat{\rho}_t(D_1) = \Tilde{\mathcal{L}}_0(D_1)\hat{\rho}_t(D_1) + \mathcal{C}\hat{\rho}_t(D_1) + \Tilde{\Gamma}\mathcal{D}[\hat{A}_1]\hat{\rho}_t(D_1) + \mathcal{F}_1(D_1)\hat{\rho}_t(D_1),
\end{equation}
where $\mathcal{F}_1(D_1)$ is given in Eq.~(\ref{eq:AppendixFokkePlanckSupOp}), and we introduced $\mathcal{C}\hat{\rho} = -ig[\dyad{L}{R}+\dyad{R}{L},\hat{\rho}]$, as well as
\begin{equation}
    \Tilde{\mathcal{L}}_0(D_1) = \mathcal{L}_1\mathcal{V}_{00} + \{ [1-\theta(D_1)]\mathcal{L}_2^{(0)} + \theta(D_1)\mathcal{L}_3^{(0)} \}\left(1-\mathcal{V}_{00} \right),
\end{equation}
with
\begin{equation}
\begin{cases}
\mathcal{L}_2^{(0)}\hat{\rho} = -[\hat{H}_0(\epsilon_d,\epsilon_d),\hat{\rho}] + \gamma_L(\epsilon_d)\mathcal{D}[\hat{\sigma}_L^\dagger] + \kappa_L(\epsilon_d)\mathcal{D}[\hat{\sigma}_L] + \gamma_R(\epsilon_d)\mathcal{D}[\hat{\sigma}_R^\dagger] + \kappa_R(\epsilon_d)\mathcal{D}[\hat{\sigma}_R], \\
\mathcal{L}_3^{(0)}\hat{\rho} = -[\hat{H}_0(\epsilon_u,\epsilon_0),\hat{\rho}] + \gamma_L(\epsilon_u)\mathcal{D}[\hat{\sigma}_L^\dagger] + \kappa_L(\epsilon_u)\mathcal{D}[\hat{\sigma}_L] + \gamma_R(\epsilon_0)\mathcal{D}[\hat{\sigma}_R^\dagger] + \kappa_R(\epsilon_0)\mathcal{D}[\hat{\sigma}_R],
\end{cases}
\end{equation}
where $\hat{H}_0(\epsilon_L,\epsilon_R) = \epsilon_L \dyad{L} + \epsilon_R \dyad{R}$ and $\hat{\sigma}_\alpha=\dyad{0}{\alpha}$.

To find the classical rate equation (\ref{eq:ClassicalQFPME}), we follow the method of Ref.~\cite{Mitchison-New-J-Phys-2015} (see Appendix C) and introduce the following Nakajima-Zwanzig superoperators \cite{Nakajima-PTP-1958,Zwanzig-JCP-1960} 
\begin{equation}
    \mathcal{P}\hat{\rho} \equiv \sum_{j=0,L,R} \mel{j}{\hat{\rho}}{j}\dyad{j}, \hspace{3cm} \mathcal{Q}\equiv 1-\mathcal{P},
\end{equation}
which single out the diagonal ($\mathcal{P}$) and off-diagonal ($\mathcal{Q}$) elements of a density matrix $\hat{\rho}$ in the basis $\{\ket{0},\ket{L},\ket{R}\}$. Both superoperators satisfy the projection property $\mathcal{P}^2=\mathcal{P}$ and $\mathcal{Q}^2=\mathcal{Q}$. Additionally note the following relations
\begin{equation}
    \begin{aligned}
        [\mathcal{L}_0(D_1),\mathcal{P}] = [\mathcal{L}_0(D_1),\mathcal{Q}] &= [\mathcal{F}_1(D_1),\mathcal{P}] =[\mathcal{F}_1(D_1),\mathcal{Q}] = [\mathcal{D}[\hat{A}_1],\mathcal{Q}] =0 \\ 
        &\mathcal{P}\mathcal{D}[\hat{A}_1]=\mathcal{D}[\hat{A}_1]\mathcal{P}=0, \\
        &\mathcal{PCP}=\mathcal{QCQ} = 0.
    \end{aligned}
    \label{eq:AppendixQTCRelations}
\end{equation}
By using these relations, we find the following coupled differential equations,
\begin{equation}
    \begin{aligned}
        \partial_t\mathcal{P}\hat{\rho}_t(D_1) &= \mathcal{L}_0\mathcal{P}\hat{\rho}_t(D_1) + \mathcal{PCQ}\hat{\rho}_t(D_1) + \mathcal{F}_1(D_1)\mathcal{P}\hat{\rho}_t(D_1), \\
        \partial_t\mathcal{Q}\hat{\rho}_t(D_1) &= \mathcal{L}_0(D_1)\mathcal{Q}\hat{\rho}_t(D_1) + \mathcal{QCP}\hat{\rho}_t(D_1) + \Tilde{\Gamma}\mathcal{D}[\hat{A}_1]\mathcal{Q}\hat{\rho}_t(D_1) + \mathcal{F}_1(D_1)\mathcal{Q}\hat{\rho}_t(D_1).
    \end{aligned}
    \label{eq:AppendixEqForPAndQ}
\end{equation}
The second of these equations has the solution
\begin{equation}
    \mathcal{Q}\hat{\rho}_t(D_1) = e^{\{\mathcal{L}_0(D_1)+\Tilde{\Gamma}\mathcal{D}[\hat{A}_1]+\mathcal{F}_1(D_1)\}(t-t_0)}\mathcal{Q}\hat{\rho}_{t_0}(D_1) + \int_{t_0}^t ds e^{\{\mathcal{L}_0(D_1)+\Tilde{\Gamma}\mathcal{D}[\hat{A}_1]+\mathcal{F}_1(D_1)\}(t-s)}\mathcal{QCP}\hat{\rho}_s(D_1).
\end{equation}
To make progress, we assume that the inverse timescale of $\mathcal{F}_1(D_1)$, $\gamma_1$, is much smaller than the dominating inverse timescale of $\mathcal{L}_0(D_1)+\Tilde{\Gamma}\mathcal{D}[\hat{A}_1]$, i.e., $\text{max}\{\Gamma,g,|\epsilon_{u/d}-\epsilon_0|,\Tilde{\Gamma}\}$, such that we can replace $\mathcal{L}_0(D_1)+\Tilde{\Gamma}\mathcal{D}[\hat{A}_1]+\mathcal{F}_1(D_1) \to \mathcal{L}_0(D_1)+\Tilde{\Gamma}\mathcal{D}[\hat{A}_1] $ in the exponential operators. We are mainly interested in the long time limit, where the first term of $\mathcal{Q}\hat{\rho}_t(D_1)$ drops out, as the real part of $\mathcal{L}_0+\Tilde{\Gamma}\mathcal{D}[\hat{A}_1]$ is negative in $\mathcal{Q}$-space. Alternatively, this term drops out by preparing the system in a diagonal state at $t_0$. In any case, the first term will be of no relevance at long times. Additionally, we make the change of variables $\tau=t-s$, and assume that the density matrix $\mathcal{P}\hat{\rho}_t(D_1)$ remains constant while the exponential factor under the integral drops to zero (Markov assumption). Under these approximations, we can write
\begin{equation}
    \mathcal{Q}\hat{\rho}_t(D_1) =  \int_{0}^\infty d\tau e^{\{\mathcal{L}_0(D_1)+\Tilde{\Gamma}\mathcal{D}[\hat{A}_1]\}\tau}\mathcal{QCP}\hat{\rho}_t(D_1) = -\{\mathcal{L}_0(D_1)+\Tilde{\Gamma}\mathcal{D}[\hat{A}_1]\}^{-1}\mathcal{CP}\hat{\rho}_t(D_1),
\end{equation}
where we introduced the Drazin inverse \cite{Mandal-JSMTE-2016,Scandi-Quantum-2019}
\begin{equation}
    \int_{0}^\infty d\tau e^{\{\mathcal{L}_0(D_1)+\Tilde{\Gamma}\mathcal{D}[\hat{A}_1]\}\tau}\mathcal{Q}= -\{\mathcal{L}_0(D_1)+\Tilde{\Gamma}\mathcal{D}[\hat{A}_1]\}^{-1},
\end{equation}
which is possible to compute as $\mathcal{L}_0(D_1)+\Tilde{\Gamma}\mathcal{D}[\hat{A}_1]$ only has nonzero eigenvalues in $\mathcal{Q}$-space. To see this, it is useful to apply the property $\mathcal{Q}^2=\mathcal{Q}$. By plugging this back into the equation for $\mathcal{P}\hat{\rho}_t(D_1)$ (\ref{eq:AppendixEqForPAndQ}), we find the following Markovian master equation
\begin{equation}
\begin{aligned}
    \partial_t\mathcal{P}\hat{\rho}_t(D_1) &= \mathcal{L}_0(D_1)\mathcal{P}\hat{\rho}_t(D_1) - \mathcal{PC}\{\mathcal{L}_0(D_1)+\Tilde{\Gamma}\mathcal{D}[\hat{A}_1]\}^{-1}\mathcal{CP}\hat{\rho}_t(D_1) + \mathcal{F}_1(D_1)\mathcal{P}\hat{\rho}_t(D_1) \\
    &= \mathcal{L}_{\rm eff}\mathcal{P}\hat{\rho}_t(D_1) + \mathcal{F}_1(D_1)\mathcal{P}\hat{\rho}_t(D_1),
\end{aligned}
\label{eq:AppendixClassicalQFPME}
\end{equation}
where
\begin{equation}
\begin{aligned}
    \mathcal{L}_{\rm eff} &= \mathcal{P}\mathcal{L}_1\mathcal{V}_{00}\mathcal{P} + [1-\theta(D_1)]\mathcal{PL}_2(1-\mathcal{V}_{00})\mathcal{P} + \theta(D_1)\mathcal{PL}_3(1-\mathcal{V}_{00})\mathcal{P} \\
    &\hspace{3cm}+\left([1-\theta(D_1)]\xi_2+\theta(D_1)\xi_3 \right)\mathcal{P}\left(\mathcal{D}[\dyad{R}{L}]+\mathcal{D}[\dyad{L}{R}] \right)\mathcal{P},
\end{aligned}
\end{equation}
with the classical rates

\begin{equation}
    \xi_2 = \frac{4g^2}{\kappa_L(\epsilon_d)+\kappa_R(\epsilon_d)+4\Tilde{\Gamma}}, \hspace{2cm} \xi_3 = \frac{4g^2[\kappa_L(\epsilon_u)+\kappa_R(\epsilon_0)+4\Tilde{\Gamma}]}{[\kappa_L(\epsilon_u)+\kappa_R(\epsilon_0)+4\Tilde{\Gamma}]^2+4(\epsilon_u-\epsilon_0)^2},
\end{equation}
for configurations 2 and 3. We thus see that the density matrix $\mathcal{P}\hat{\rho}_t(D_1)$ in Eq.~(\ref{eq:AppendixClassicalQFPME}) changes on a timescale determined by the rates $\Gamma$, $\xi_2$, $\xi_3$, and $\gamma_1$. To justify the approximations leading to Eq.~(\ref{eq:AppendixClassicalQFPME}), we require that $\Tilde{\Gamma}\gg\Gamma,\xi_2,\xi_3,\gamma_1$. By vectorizing $\mathcal{P}\hat{\rho}_t(D_1)\to[\rho_{00}(D_1),\rho_{LL}(D_1),\rho_{RR}(D_1)]^{\rm T}\equiv\boldsymbol{\rho}_t(D_1)$, where $\rho_{aa}(D_1)=\mel{a}{\hat{\rho}_t(D_1)}{a}$, as done in the main text, Eq.~(\ref{eq:AppendixClassicalQFPME}) can be written as the matrix equation (\ref{eq:ClassicalQFPME}).

We can now use Eq.~(\ref{eq:AppendixClassicalQFPME}) to derive Eq.~(\ref{eq:ClassicalIdealModel}). We begin by assuming that detector 1 is fast, i.e., $\gamma_1\gg\Gamma,\xi_2,\xi_3$. Under this hierarchy of timescales, we can use the methods of Appendix \ref{appendix:FastDetectors} and write the density matrix, to zeroth order in $1/\gamma_1$, as
\begin{equation}
    \mathcal{P}\hat{\rho}_t(D_1) = \sum_{a=0,L,R} \pi_{aa}^{(1)}(D_1)\mathcal{V}_{aa}\mathcal{P}\hat{\rho}_t,
\end{equation}
where $\pi_{aa}^{(1)}(D_1)$ is defined in Eq.~(\ref{eq:AppendixDetector}), $\mathcal{V}_{aa}$ is defined below Eq.~(\ref{eq:AppendixZerothOrderDensityMatrixD2}), and $\mathcal{P}\hat{\rho}_t=\int dD_1 \mathcal{P}\hat{\rho}_t(D_1)$. With this representation of $\mathcal{P}\hat{\rho}_t(D_1)$, $\mathcal{F}_1(D_1)\mathcal{P}\hat{\rho}_t(D_1)=0$. Integrating Eq.~(\ref{eq:AppendixClassicalQFPME}) over $D_1$ gives that
\begin{equation}
    \begin{aligned}
        &\int dD \mathcal{L}_{\rm eff}(D_1)\mathcal{P}\hat{\rho}_t(D_1) \\
        &= \Bigg\{ \mathcal{P}\mathcal{L}_1\mathcal{V}_{00}\mathcal{P} + \mathcal{P}\mathcal{L}_2\Big[(1-\eta_1)\mathcal{V}_{LL}+\eta_1\mathcal{V}_{RR} \Big]\mathcal{P} + \mathcal{P}\mathcal{L}_3\Big[\eta_1\mathcal{V}_{LL}+(1-\eta_1)\mathcal{V}_{RR} \Big]\mathcal{P} \\
        &+\mathcal{P}\Big(\mathcal{D}[\dyad{L}{R}]+\mathcal{D}[\dyad{R}{L}]\Big)\mathcal{P}\Big(\xi_2\Big[(1-\eta_1)\mathcal{V}_{LL}+\eta_1\mathcal{V}_{RR}\Big] + \xi_3\Big[\eta_1\mathcal{V}_{LL}+(1-\eta_1)\mathcal{V}_{RR}\Big] \Big)\Bigg\}\mathcal{P}\hat{\rho}_t \equiv \mathcal{L}_{\rm cl}\mathcal{P}\hat{\rho}_t,
    \end{aligned}
    \label{eq:AppendixSepTSClassicalModel}
\end{equation}
where $\eta_1$ is the feedback error probability defined in Eq.~(\ref{eq:AppendixErrorProb}). By vectorizing the density matrix $\mathcal{P}\hat{\rho}_t$ as in the main text, $\mathcal{L}_{\rm cl}$ can be written as a rate matrix as
\begin{equation}
    \mathcal{L}_{\rm cl} \to
    \begin{pmatrix}
     -\gamma_L(\epsilon_0)-\gamma_R(\epsilon_u) & (1-\eta_1)\kappa_L(\epsilon_d)+\eta_1\kappa_L(\epsilon_u) & (1-\eta_1)\kappa_R(\epsilon_0)+\eta_1\kappa_R(\epsilon_d) \\
     \gamma_L(\epsilon_0) & x & (1-\eta_1)\xi_3+\eta_1\xi_2 \\
     \gamma_R(\epsilon_u) & (1-\eta_1)\xi_2+\eta_1\xi_3 & y
    \end{pmatrix},
    \label{eq:AppendixMatrixRepresentationClassicalModel}
\end{equation}
where $x=-(1-\eta_1)\kappa_L(\epsilon_d)-\eta_1\kappa_L(\epsilon_u) - (1-\eta_1)\xi_2-\eta_1\xi_3$ and $y=-(1-\eta_1)\kappa_R(\epsilon_0)-\eta_1\kappa_R(\epsilon_d)-(1-\eta_1)\xi_3-\eta_1\xi_2$. For strong measurements $\lambda_1\gg\gamma_1$, $\eta_1\to0$ exponentially, while $\xi_{2/3}$ go to zero as $1/\lambda_1$. By additionally assuming that $|\epsilon_u-\epsilon_0|$ is large and $|\epsilon_{u/d}-\mu_{L/R}|\gg T$, $\xi_3\to0$ and $\gamma_{L/R}(\epsilon_u),\kappa_{L/R}(\epsilon_d)\to0$, leaving us with
\begin{equation}
    \begin{aligned}
        \mathcal{L}_{\rm cl} = \mathcal{P}\mathcal{L}_1\mathcal{V}_{00}\mathcal{P} + \mathcal{P}\mathcal{L}_2\mathcal{V}_{LL}\mathcal{P} + \mathcal{P}\mathcal{L}_3\mathcal{V}_{RR}\mathcal{P}
        +\xi_2\mathcal{P}\Big(\mathcal{D}[\dyad{L}{R}]+\mathcal{D}[\dyad{R}{L}]\Big)\mathcal{V}_{LL}\mathcal{P}.
    \end{aligned}
\end{equation}
Vectorizing $\mathcal{P}\hat{\rho}_t$, similarly as above, we find the matrix equation (\ref{eq:ClassicalIdealModel}).

\section{Experimentally accessible observables}
\label{appendix:PostProcess}
The observables $\hat{A}_1$ and $\hat{A}_2$ are idealizations of what the detectors in an actual experiment would measure. They are nevertheless useful to study as they simplify many of the calculations and provide valuable insights into the performance of the demon. In an actual experiment, the detectors would measure the observables $\hat{A}=a_0\dyad{0}+a_L\dyad{L}+a_R\dyad{R}$, and $\hat{B}=b_0\dyad{0}+b_L\dyad{L}+b_R\dyad{R}$, where the real numbers $a_j$ and $b_j$ ($j=0,L,R$) tell us how the detectors couple to the charge states of the system. Here, we treat these numbers as arbitrary constants to represent any coupling. For the remainder of this appendix, we discuss how the QFPME (\ref{eq:FullQFPME}) is modified for $\hat{A}$ and $\hat{B}$ when using the feedback protocol defined in Eq.~(\ref{eq:FullFeedbackLiouvillian}).

We begin by noting that the outcomes $D_A$ and $D_B$ of $\hat{A}$ and $\hat{B}$ drift towards the points $(a_0,b_0)$, $(a_L,b_L)$, and $(a_R,b_R)$ for the charge states $\dyad{0}$, $\dyad{L}$, and $\dyad{R}$. These points form a triangle in the two-dimensional outcome space spanned by $D_A$ and $D_B$. By dividing this space into three regions, we can define a feedback protocol for the demon, similar to the one illustrated in Fig.~\ref{fig:System}(c). In general, the division of these regions will complicate the analysis of the problem. For the protocol in Fig.~\ref{fig:System}(c) this was not a problem, as the regions are separated by lines along the $x$ and $y$ axes. To this end, we want to post-process $D_A$ and $D_B$ such that the processed outcomes drift towards the points $(0,1)$, $(-1,-1)$, and $(1,-1)$, just as for the measurements of $\hat{A}_1$ and $\hat{A}_2$. This allows us to use the protocol illustrated in Fig.~\ref{fig:System}(c), simplifying analytical calculations. For the remainder of this appendix, we show how this is done.

Our starting point is the QFPME for measuring $\hat{A}$ and $\hat{B}$, which is given by
\begin{equation}
\begin{aligned}
      \partial_t&\hat{\rho}_t(\boldsymbol{D}) = \mathcal{L}(\Tilde{\boldsymbol{D}})\hat{\rho}_t(\boldsymbol{D}) + \tilde{\lambda}\mathcal{D}[\hat{A}]\hat{\rho}_t(\boldsymbol{D}) \\
      &+ \frac{\gamma_A}{2}\partial_{D_A}\{D_A-\hat{A},\hat{\rho}_t(\boldsymbol{D})\} + \frac{\gamma_A^2}{8\lambda_A}\partial_{D_A}^2\hat{\rho}_t(\boldsymbol{D}) \\
      &+ \frac{\gamma_B}{2}\partial_{D_B}\{D_B-\hat{B},\hat{\rho}_t(\boldsymbol{D})\} + \frac{\gamma_B^2}{8\lambda_B}\partial_{D_B}^2\hat{\rho}_t(\boldsymbol{D}),
\end{aligned}
\label{eq:appendix:QFPMEAB}
\end{equation}
where $\boldsymbol{D}=(D_A,D_B)$, $\tilde{\lambda}=\lambda_A+\lambda_B(b_L-b_R)^2/(a_L-a_R)^2$ is the collective dephasing rate due to measurement backaction, with $\lambda_{A(B)}$ being the measurement strength for measuring $\hat{A}(\hat{B})$, and $\gamma_{A(B)}$ is the bandwidth of the detector measuring $\hat{A}(\hat{B})$. Note that the feedback is based on the post-processed outcomes $\Tilde{\boldsymbol{D}}=(D,D')$, where
\begin{equation}
\begin{pmatrix}
    D \\ D'
\end{pmatrix}
=
\begin{pmatrix}
    x & y \\ z & w 
\end{pmatrix}
\begin{pmatrix}
    D_A \\ D_B
\end{pmatrix}
+
\begin{pmatrix}
    D_0 \\ D_0'
\end{pmatrix},
    \label{eq:appendix:PostProcessing}
\end{equation}
with the real numbers $x$, $y$, $z$, $w$, $D_0$, and $D_0'$. By choosing these numbers as
\begin{equation}
    \begin{aligned}
        x &= (b_L+b_R-2b_0)/\mathcal{N}, \\
        y &= (2a_0-a_L-a_R)/\mathcal{N}, \\
        z &= 2(b_R-b_L)/\mathcal{N}, \\
        w &= 2(a_L-a_R)/\mathcal{N},\\
        D_0 &= \left[ b_0(a_L+a_R)-a_0(b_L+b_R)\right]/\mathcal{N}, \\
        D_0' &= \left[a_0(b_L-b_R)-a_L(b_0+b_R)+a_R(b_0+b_L) \right]/\mathcal{N}, \\
        \mathcal{N} &= a_0(b_R-b_L) + a_L(b_0-b_R) + a_R(b_L-b_0),
    \end{aligned}
\end{equation}
we make sure that $\tilde{\boldsymbol{D}}$ drifts towards the points $(0,1)$, $(-1,-1)$ and $(1,-1)$ for the charge states $\dyad{0}$, $\dyad{L}$, and $\dyad{R}$, respectively. In this way, we can use the feedback protocol illustrated in Fig.~\ref{fig:System}(c). Under the transformation in Eq.~(\ref{eq:appendix:PostProcessing}), we find, with the chain rule, $\partial_{D_A} = x\partial_{D} + z\partial_{D'} $ and $\partial_{D_B}=y\partial_{D}+w\partial_{D'}$. Using this together with $\tilde{\rho}_t(\tilde{\boldsymbol{D}})=\int dD_AdD_B \delta[D-(xD_A+yD_B)]\delta[D'-(zD_A+wD_B)]\hat{\rho}_t(\boldsymbol{D})$, Eq.~(\ref{eq:appendix:QFPMEAB}) transforms into
\begin{equation}
\begin{aligned}
        \partial_t\tilde{\rho}_t(\tilde{\boldsymbol{D}}) &= \mathcal{L}(\Tilde{\boldsymbol{D}})\tilde{\rho}_t(\tilde{\boldsymbol{D}}) + \Tilde{\lambda}\mathcal{D}[\hat{A}]\tilde{\rho}_t(\tilde{\boldsymbol{D}})  \\
    &+ \frac{1}{2}\partial_D\left\{ (\gamma_A x \alpha+\gamma_B y \delta)(D-D_0) + (\gamma_A x \beta + \gamma_B y \xi )(D'-D_0') - \left(\gamma_A x \hat{A} + \gamma_B y \hat{B} \right) , \hat{\rho}_t(\boldsymbol{D}) \right\} \\
    &+ \frac{1}{2}\partial_{D'}\left\{ (\gamma_A z \alpha+\gamma_B w \delta)(D-D_0) + (\gamma_A z \beta + \gamma_B w \xi )(D'-D_0') - \left(\gamma_A z \hat{A} + \gamma_B w \hat{B} \right) , \hat{\rho}_t(\boldsymbol{D}) \right\} \\
    &+ \left( \frac{x^2\gamma_A^2}{8\lambda_A}+\frac{y^2\gamma_B^2}{8\lambda_B} \right)\partial_D^2\tilde{\rho}_t(\tilde{\boldsymbol{D}}) + \left( \frac{z^2\gamma_A^2}{8\lambda_A}+\frac{w^2\gamma_B^2}{8\lambda_B} \right)\partial_{D'}^2\tilde{\rho}_t(\tilde{\boldsymbol{D}}) + \left(\frac{xz\gamma_A^2}{4\lambda_A}+\frac{yw\gamma_B^2}{4\lambda_B}\right)\partial_D\partial_{D'}\tilde{\rho}_t(\tilde{\boldsymbol{D}}),
\end{aligned}
\end{equation}
where $\alpha=w/(xw-yz)=(a_R-a_L)/2$, $\beta=-y/(xw-yz)=(2a_0-a_L-a_R)/4$, $\delta=-z/(xw-yz)=(b_R-b_L)/2$, and $\xi=x/(xw-yz)=(2b_0-b_L-b_R)/4$ are obtained by inverting the matrix in Eq.~(\ref{eq:appendix:PostProcessing}). By mixing the signals $D_A$ and $D_B$, the noise of $D$ and $D'$ gets correlated, as illustrated by the last term of the third row, and the drift terms become dependent on the other measurement outcome.

\section{Stochastic simulation of quantum trajectories}
\label{appendix:MCWF}

In this appendix we describe the simulation scheme used to obtain the trajectories shown in Fig.~\ref{fig:performance_simplifying_assumptions} and Fig.~\ref{fig:QtoCTrajectories}. Here we implement the quantum jump trajectory \cite{goan-2001-PRB-quantum_traj,Wiseman-Milburn-PRA-1993-quantum_traj} simulation under feedback control by modifying the standard Monte Carlo Wave Function (MCWF) \cite{Molmer:93,Daley-2014-AIP-review_mcwf} algorithm to incorporate measurement and feedback. To study the model under consideration, we implement a simulation scheme that captures the dynamics of the detector and the quantum state together.  At any time instant $t$, we describe the quantum state of the DQD system by the wave function:
\begin{equation}
\ket{\psi(t)}=c_0(t)\ket{0}+c_L(t)\ket{L}+c_R(t)\ket{R} 
\end{equation}
Here the coefficients $\{c_j(t)\}_{j=L,R,E}$ depend on the complete history of the measurement outcomes.
 The conditional density matrix corresponding to this quantum state is by $\hat{\rho}_c(t)=\ket{\psi(t)}\bra{\psi(t)}$.
 \subsection{Evolution map}
 For our simulations, we model the stochastic evolution of the quantum state $\hat{\rho}_c(t)$  under measurement and feedback by the evolution map,
\begin{equation}
    \hat{\rho}_c(t+dt) = \left( \sum_{j=1}^4 dN_j(t)\mathcal{E}_{\rm J}^{(j)}(\boldsymbol{D}) + dN_0(t)\mathcal{E}_{\rm NJ}(\boldsymbol{D}) \right)\mathcal{M}_2(z_2)\mathcal{M}_1(z_1)\hat{\rho}_c(t).
\label{app_eq:trajectory_update}
\end{equation}
Here $\mathcal{M}_1(z_1)$ and $\mathcal{M}_2(z_2)$ are measurement superoperators corresponding to detector 1 and detector 2, respectively, and these are given by:
\begin{equation}
    \label{app_eq:measurement_superoperator}\mathcal{M}_l(z)\hat{\rho}=\frac{\hat{K}_l(z)\hat{\rho}\hat{K}_l^\dagger(z)}{\trace\{\hat{K}_l^\dagger(z)\hat{K}_l(z)\hat{\rho}\}},\ \  \hat{K}_l(z) = \left( \frac{2\lambda_ldt}{\pi} \right)^{1/4} e^{-\lambda_l dt\left( z-\hat{A}_l\right)^2}, \ \ l=1,2.
\end{equation}
The  distribution of the measurement outcomes $z_l(t)$ are given by $P_l^{(t)}(z)=\trace\{\hat{K}_l^\dagger(z)\hat{K}_l(z)\hat{\rho}_c(t)\}$.
The superoperators, $\mathcal{E}^{(j)}_{\mathrm{J}}(\boldsymbol{D})$ and $\mathcal{E}_{\mathrm{NJ}}(\boldsymbol{D})$ correspond to `Jump' and `No Jump' evolution of the quantum-jump unravelling of the Lindblad master equation given in Eq.~\eqref{eq:DQDDynamics}, where the energy levels $(\epsilon_L(\boldsymbol{D}),\epsilon_R(\boldsymbol{D}))$ are now parameterized by the control parameters $\boldsymbol{D}$, making them feedback dependent. In the evolution map given in Eq.~\eqref{app_eq:trajectory_update}, the random variables $\{z_l\}$ capture the stochasticity due to the measurement process and the random variables $\{dN_j\}$ capture the stochasticity due to the interaction with the the electron reservoirs. At any instant, the value of the control parameter $\boldsymbol{D}(t)=(D_1(t),D_2(t))$ are calculated from the past measurement outcomes $(z_1(t),z_2(t))$ through the filtering relation  
\begin{equation}
    \label{app_eq:D_integrated_equation}
    D_j(t) = \int_{-\infty}^t ds \gamma_j e^{-\gamma_j(t-s)}z_j(s), \ \ j=1,2.
\end{equation}
 The unravelling of the Lindblad term in Eq.~\eqref{eq:DQDDynamics} leads to four possible quantum jumps which we describe by the jump operators: $\hat{c}_1(\boldsymbol{D})=\sqrt{ \gamma_L(\epsilon_L(\boldsymbol{D}))} |L\rangle\langle 0|$, $\hat{c}_2(\boldsymbol{D})=\sqrt{ \kappa_L(\epsilon_L(\boldsymbol{D}))} |0\rangle\langle L|$, $\hat{c}_3(\boldsymbol{D})=\sqrt{ \gamma_R(\epsilon_R(\boldsymbol{D}))} |R\rangle\langle 0|$ and $\hat{c}_4(\boldsymbol{D})=\sqrt{ \kappa_R(\epsilon_R(\boldsymbol{D}))} |0\rangle\langle R|$. The corresponding jump superoperators are given as
\begin{equation}
    \mathcal{E}_{\rm J}^{(j)}(\boldsymbol{D})\hat{\tilde{\rho}}_c = \frac{\hat{c}_j(\boldsymbol{D})\hat{\tilde{\rho}}_c\hat{c}_j^\dagger(\boldsymbol{D})}{\trace\{\hat{c}_j^\dagger(\boldsymbol{D})\hat{c}_j(\boldsymbol{D})\hat{\tilde{\rho}}_c\}}, \ \ j=1,..,4, 
\end{equation}
where the state of the DQD is $\hat{\tilde{\rho}}_c(t)$ at time instant $t$. Here we use $\hat{\tilde{\rho}}_c(t)$ instead of $\hat{\rho}_c$ to imply that the evolution superoperators are being applied on the conditional density matrix $\hat{\tilde{\rho}}_c(t)$ which is obtained from $\hat{\rho}_c$ by applying the measurement superoperators.  The probability of a particular jump $j$ happening during the time interval $t$ to $t+dt$ is given by $p_j(t)=dt\trace\{\hat{c}_j^\dagger(\boldsymbol{D})\hat{c}_j(\boldsymbol{D})\hat{\tilde{\rho}}_c\}$ and the stochastic jump random variables $\{dN_j(t), j =1,..,4\}$ are distributed as
\begin{equation}
    \label{app_eq:dNj}
    dN_j(t)= \begin{cases}
        1, \ \ \mathrm{Pr}[1] =p_j(t), \\
        0, \ \ \mathrm{Pr}[0] =1 - p_j(t),
    \end{cases}
\end{equation}
where $\text{Pr}[0(1)]$ denotes the probability of observing $dN_j(t)=0(1)$. The stochastic jump variables satisfy $\text{E}[dN_j(t)]=p_j(t)$, and $dN_i(t)dN_j(t) = \delta_{ij} dN_{j}(t)$, with $\text{E}[\cdot]$ being the ensemble average over all possible trajectories of jumps. We also define a similar stochastic variable
\begin{equation}
    \label{app_eq:dN0}
    dN_0(t)=1-\sum_{j=1}^4 dN_j(t)
\end{equation}
for the `No Jump' evolution during the interval $ [t,t+dt] $. For $dN_0(t)$, we have the expectation value $\text{E}[dN_0(t)]=p_0(t)=1-\sum_jp_j(t)$. For the `No Jump' case, the quantum state evolves under the effective non-unitary Hamiltonian,
\begin{equation}
\label{app_eq:H_eff}
    \hat{H}_{\rm eff}(\boldsymbol{D})=\hat{H}(\epsilon_L(\boldsymbol{D}),\epsilon_R(\boldsymbol{D}))-\frac{i}{2}\sum_{j=1}^4\hat{c}_j^\dagger(\boldsymbol{D})\hat{c}_j(\boldsymbol{D}),
\end{equation}
where, $\hat{H}(\epsilon_L(\boldsymbol{D}),\epsilon_R(\boldsymbol{D}))$ is the feedback Hamiltonian as defined in the Eq.~\eqref{eq:DQDHamiltonian}. The No-Jump (NJ) evolution superoperator is then defined as,
\begin{equation}
\label{app_eq:NoJumpMap}
    \mathcal{E}_{\rm NJ}(\boldsymbol{D})\hat{\tilde{\rho}}_c = \frac{e^{-i\hat{H}_{\rm eff}(\boldsymbol{D})dt}\hat{\tilde{\rho}}_c e^{i\hat{H}_{\rm eff}^\dagger(\boldsymbol{D})dt}}{\trace\{{e^{-i\hat{H}_{\rm eff}(\boldsymbol{D})dt}\hat{\tilde{\rho}}_c e^{i\hat{H}_{\rm eff}^\dagger(\boldsymbol{D})dt}\}}}.
\end{equation}

\subsection{Ideal charge detection assumption}
We express the filtering in Eq.~\eqref{app_eq:D_integrated_equation} as a difference equation by discretizing time $t=k\delta t$ as
\begin{equation}
    D_j(k \delta t) = \delta t \gamma_j z_j(k \delta t) + (1 - \delta t \gamma_j) D_j((k-1) \delta t)
\end{equation}
where $\delta t$ is the time-step of the simulation. Here $(\gamma_j \delta t)$ is the smoothing factor of the filter and $\delta t$ needs to be chosen such that we have $ (\gamma_j \delta t) \leq 1$.
Now, similar to the approximations made in the Appendix. \ref{appendix:FastDetectors}, we assume $\gamma_2\gg\text{max}\{\Gamma,g,|\epsilon_{u/d}-\epsilon_0|,\Tilde{\Gamma},\gamma_1\}$. To achieve the fast limit of the detector we take the maximum possible value of $\gamma_2$ in the simulation by setting $\gamma_2 \delta t =1$. Thus for the fast detector 2 we have,
\begin{equation}
    D_2(k \delta t) =z_2(k \delta t).
\end{equation}
For detector 2, (observable: $\hat{A}_2= |0\rangle\langle 0 | -( |L\rangle\langle L |+ |R\rangle\langle R |)$ we have the POVM $\hat{K}_2(z)$ in the measurement basis as
\begin{equation}
   \hat{K}_2(z_2) =  \left( \frac{2\lambda_2 \delta t}{\pi} \right)^{1/4}  \left[e^{-\lambda_2  \delta t\left( z_2-1\right)^2} |0\rangle\langle 0 | + e^{-\lambda_2  \delta t\left( z_2+1)\right)^2} (|L\rangle\langle L |+ |R\rangle\langle R |)\right].
\end{equation}
Next we want to increase the measurement strength of the detector 2 to the infinite $(\lambda_2 \to \infty)$ while keeping $\delta t$ fixed to achieve the strong measurement limit in the POVM $\hat{K}_2(z_2)$, which we can approximate as projective measurement POVM: 
\begin{equation}
    \hat{K}_2(z_2)=\begin{cases}
    |0\rangle\langle0|, z_2= 1, \\
    (\mathbb{1}-|0\rangle\langle0|), z_2= -1 \\
    0, \mathrm{otherwise}
    \end{cases}
\end{equation}
and for a quantum state $\hat{\rho}_c(t)$, the corresponding distribution of $z_2$ is given as 
$P_2^{(t)}(z_2 = 1)= \trace\{|0\rangle\langle 0|\hat{ \rho}_c(t)\}$ and $P_2^{(t)}(z_2 = -1)=1 -\trace\{|0\rangle\langle 0|\hat{\rho}_c(t)\}$.

\subsection{Implementation}
At any time instant $t$, we have either $c_0(t)=1, c_L(t)=c_R(t)=0$ or, $c_0(t)=0, |c_L(t)|^2 + |c_R(t)|^2 =1$.
The quantum state vector $\ket{\psi(t)}$ is represented as a $3 \times 1$ normalized column vector. The control parameter starts with unupdated state at $\boldsymbol{D}(t - \delta t) =(D_1(t - \delta t), D_2(t - \delta t))$.\par 

\underline{Detector-1 Measurement:} To simulate the effect of the measurement, first a random variable $\xi \in \{-1,0,+1\}$ is sampled based on the distribution
\begin{equation}
    \mathrm{Pr}(\xi) =\begin{cases}
       |c_L(t)|^2 , \xi  = -1 \\
       |c_0(t)|^2 , \xi  = 0 \\
       |c_R(t)|^2 , \xi  = +1.
    \end{cases}
\end{equation}
Then the measurement outcome is sampled from the normal distribution with $\xi$ mean ($\mu$) and $1/(2\sqrt{\lambda_1 \delta t})$ standard deviation ($\sigma$): 
\begin{equation}
    z_1(t) \sim\mathcal{N}\left(\mu=\xi,\sigma^2=\frac{1}{4 \lambda_1 \delta t}\right).
\end{equation}
Then using the sampled value of $z_1(t)$, the matrix representation of $\hat{K_1}(z_1(t))$ is calculated. We then modify the quantum state $\ket{\psi(t)}$  to a intermediate new state as
\begin{equation}
    \ket{\Tilde{\psi}(t)} =\frac{\hat{K_1}(z_1(t)) \ket{\psi(t)}}{\left|\hat{K_1}(z_1(t)) \ket{\psi(t)}\right|}.
\end{equation}
Next the updated control parameter $D_1(t)$ is calculated using the measurement outcome $z_1(t)$ as:
\begin{equation}
     D_1( t) = \delta t \gamma_1 z_1(t) + (1 - \delta t \gamma_1) D_1(t-\delta t).
\end{equation}
\underline{Detector 2 Measurement}:
The second detector performs projective measurements in the system. Thus, we have  $z_2(t) \in \{-1,1\}$ and the distribution of $z_2(t)$ is given as
\begin{equation}
    \mathrm{Pr}(z_2(t)) = \begin{cases}
         |\langle 0 |\Tilde{\psi}(t) \rangle|^2, \ z_2(t) =1, \\
         1- |\langle 0 |\Tilde{\psi}(t) \rangle|^2, \ z_2(t) =-1.
    \end{cases}
\end{equation}
Now the updated control parameter is given as
\begin{equation}
    D_2(t)=z_2(t),
\end{equation}
since we are working at the ideal detector limit where $\gamma_2 \to \infty$. The post measurement state after detector 2 measurement $\ket{\psi'(t)}$ is given as 
\begin{equation}
    \ket{\psi'(t)}=\begin{cases}
        \ket{0}, \ z_2(t)=1, \\
         (\mathbb{1}-|0\rangle\langle0|)|\tilde{\psi}(t)\rangle, z_2(t) =-1.
    \end{cases}
\end{equation}
Since we are working in the ideal charge detection approximation, and the quantum state $|\Tilde{\psi}(t)\rangle$ is either $\ket{0}$ or of the form $c_L(t)\ket{L}+c_R(t)\ket{R}$, the post-measurement state $\ket{\psi'(t)}$ remains unchanged from the state $|\Tilde{\psi}(t)\rangle$ under the measurement by detector 2.
\par
\underline{State Update:}

The final time evolved state $\ket{\psi(t+dt)}$ can be obtained by the stochastic evolution equation
\begin{equation}
\label{eq:StochUpdatePOVM}
    \ket{\psi(t+\delta t)}= \sum_{k=0}^{4}dN_k(t)\frac{\hat{c}_k(D_1(t),D_2(t))\ket{\psi'(t)}}{|\hat{c}_k(D_1(t),D_2(t))\ket{\psi'(t)}|},
\end{equation}
where $\hat{c}_1 = \sqrt{\gamma_L(\epsilon_L)}\dyad{L}{0}$, $\hat{c}_2 = \sqrt{\kappa_L(\epsilon_L)}\dyad{0}{L}$, $\hat{c}_3 = \sqrt{\gamma_R(\epsilon_R)}\dyad{R}{0}$, $\hat{c}_4 = \sqrt{\kappa_R(\epsilon_R)}\dyad{0}{R}$ and $\hat{c}_0=e^{-i\delta t\hat{H}_{\mathrm{eff}}}$, where, $\epsilon_{L/R}\equiv\epsilon_{L/R}(D_1(t),D_2(t))$ and the effective non-unitary Hamiltonian $\hat{H}_{\mathrm{eff}}(D_1(t),D_2(t))$ is given by Eq.~\eqref{app_eq:H_eff}. The matrix representation of the operator $\hat{c}_0$ is calculated numerically by direct exponentiation of the matrix $-i \delta t \hat{H}_{\mathrm{eff}}(D_1(t),D_2(t))$.
Each of the stochastic variables are defined by the Eqs.~\eqref{app_eq:dNj} and \eqref{app_eq:dN0}. To simulate this step, first the stochastic variable corresponding to No-Jump evolution $dN_0(t)$, is sampled as 
\begin{equation}
    \mathrm{Pr}(dN_0(t)) = \begin{cases}
        \delta t \sum_{k=1}^4 \langle \psi'(t)|\hat{c}^{\dagger}_k\hat{c}_k|\psi'(t)\rangle, \ dN_0= 0\\
        1- \delta t \sum_{k=1}^4 \langle \psi'(t)|\hat{c}^{\dagger}_k\hat{c}_k|\psi'(t)\rangle, \ dN_0=1.
    \end{cases}
\end{equation}
If the sampling results in $dN_0(t) =1$, then the updated state is calculated directly as 
\begin{equation}
    |\psi(t+\delta t)\rangle = \frac{\hat{c}_0|\psi'(t)\rangle}{|\hat{c}_0|\psi'(t)\rangle|}.
\end{equation}
If the sampling results in $dN_0(t)=0$ then another Monte-Carlo step is made to determine which jump is happening by sampling the variable $k^*$ where $k^*\in\{1,2,3,4\}$. The distribution of $k^*$ is given as,
\begin{equation}
    \mathrm{Pr}(k^*)=\frac{\langle \psi'(t)|\hat{c}^{\dagger}_{k^*}\hat{c}_{k^*}|\psi'(t)\rangle}{\sum_{k'=1}^4 \langle \psi'(t)|\hat{c}^{\dagger}_{k'}\hat{c}_{k'}|\psi'(t)\rangle}, \ k^*\in \{1,2,3,4\}. 
\end{equation}
Then the quantum state is updated by applying the corresponding jump operator to $k^*$ as,
\begin{equation}
    \psi(t+\delta t)= \frac{\hat{c}_{k^*}|\psi'(t)\rangle}{|\hat{c}_{k^*}|\psi'(t)\rangle|}.
\end{equation}

\end{document}